\documentclass[twocolumn,twocolappendix]{aastex631}
\usepackage{comment}
\usepackage[version=4]{mhchem}
\usepackage{chemformula}
\usepackage{booktabs}
\usepackage{threeparttable}
\usepackage{rotating}
\usepackage{mathtools}
\usepackage{graphicx}
\usepackage{soul}
\usepackage{hyperref}
\usepackage{natbib}

\graphicspath{{./}{figures/}}

\received{January 1, 2018}
\revised{January 7, 2018}
\accepted{\today}
\submitjournal{ApJ}

\shortauthors{Ferrero et al.}

\begin{document}

 \title{Formation of interstellar complex organic molecules on water-rich ices triggered by atomic carbon freezing}

\correspondingauthor{Albert Rimola, Mariona Sodupe}
\email{albert.rimola@uab.cat, mariona.sodupe@uab.cat}

\author[0000-0001-7819-7657]{Stefano Ferrero}
\affiliation{Departament de Qu\'{i}mica, Universitat Aut\`{o}noma de Barcelona, Bellaterra, 08193, Catalonia, Spain}

\author[0000-0001-9664-6292]{Cecilia Ceccarelli}
\affiliation{Univ. Grenoble Alpes, CNRS, Institut de Plan\'{e}tologie et d'Astrophysique de Grenoble (IPAG), 38000 Grenoble, France}

\author[0000-0001-8886-9832]{Piero Ugliengo}
\affiliation{Dipartimento di Chimica and Nanostructured Interfaces and Surfaces (NIS) Centre, Universit\`{a} degli Studi di Torino, via P. Giuria 7, 10125, Torino, Italy}

\author[0000-0003-0276-0524]{Mariona Sodupe}
\affiliation{Departament de Qu\'{i}mica, Universitat Aut\`{o}noma de Barcelona, Bellaterra, 08193, Catalonia, Spain}

\author[0000-0002-9637-4554]{Albert Rimola}
\affiliation{Departament de Qu\'{i}mica, Universitat Aut\`{o}noma de Barcelona, Bellaterra, 08193, Catalonia, Spain}

\begin{abstract}
The reactivity of interstellar carbon atoms (C) on the water-dominated ices is one of the possible ways to form interstellar complex organic molecules (iCOMs).
In this work, we report a quantum chemical study of the coupling reaction of C ($^3$P) with an icy water molecule, alongside possible subsequent reactions with the most abundant closed shell frozen species (NH$_3$, CO, CO$_2$ and H$_2$), atoms (H, N and O), and molecular radicals (OH, NH$_2$ and CH$_3$). 
We found that C spontaneously reacts with the water molecule, resulting in the formation of $^3$C-OH$_2$, a highly reactive species due to its triplet electronic state.
While reactions with the closed-shell species do not show any reactivity, reactions with N and O  form CN and CO, respectively, the latter ending up into methanol upon subsequent hydrogenation.
The reactions with OH, CH$_3$ and NH$_2$ form methanediol, ethanol and methanimine, respectively, upon subsequent hydrogenation.
We also propose an explanation for methane formation, observed in experiments through H additions to C in the presence of ices.
The astrochemical implications of this work are: 
 i) atomic C on water ice is locked into $^3$C-OH$_2$, making difficult the reactivity of bare C atoms on the icy surfaces, contrary to what is assumed in astrochemical current models; and
 ii) the extraordinary reactivity of $^3$C-OH$_2$ provides new routes towards the formation of iCOMs in a non-energetic way, in particular ethanol, mother of other iCOMs once in the gas-phase.
\end{abstract}



\section{Introduction} \label{sec:intro}

Improving our understanding of the formation of molecules at the cryogenic temperatures of the interstellar molecular clouds is of paramount importance to fully understand their chemical evolution (diversity and complexity). 
The surfaces of interstellar grains, sub-micrometre sized particles present in the molecular clouds and well mixed with the gas, are potential sites where reactions forming interstellar molecules can take place \citep[e.g.][]{Allen1975-surfacechemistry,duley1978,Tielens1982}. 
This is because grains can play different roles in these synthesis reactions: 
(i) as chemical catalysts, by reducing the activation energies \citep{watanabe_ice_2008,zamirri_quantum_2019, cuppen_grain_2017}; 
(ii) as third bodies, through which the reaction energies can be dissipated hence stabilising the newly formed species \citep{zamirri_quantum_2019,cuppen_grain_2017}; 
(iii) as reactant concentrators, which retain the reactive species and allow their diffusion on the surfaces for an eventual reactive encounter \citep{zamirri_quantum_2019, hama_surface_2013, cuppen_grain_2017, Ceccarelli2022-PP7}; and iv) in the case of ices, as reactant suppliers, since icy components can be converted into reactive derivatives ready to react (e.g., generation of radicals upon UV incidence)
\citep{Oberg2016, Ceccarelli2022-PP7}. 

In general, interstellar grains are made of a refractory core, consisting of silicate or carbonaceous materials \citep[e.g.][]{henning_cosmic_2010,jones_evolution_2013, jones_global_2017}, and, in the coldest and densest regions of the Interstellar Medium (ISM), by icy mantles coating them \citep[e.g.][]{boogert_observations_2015}. 
The grain mantles are principally made of water and other molecules (e.g., CO, CO$_2$, NH$_3$, CH$_3$OH, CH$_4$, among others) in smaller quantities, detected employing infrared (IR) spectroscopic observations \citep[e.g.][]{boogert_observations_2015, McClure2023, Yang2022-JWST}. 
An important constraint on the possible interstellar chemical reactions is set by the very low temperatures (in molecular clouds $\leq 10$ K), which only permit essentially barrierless or very low energy barrier reactions to be efficient, unless external energetic inputs are involved \citep[e.g.][]{Schutte1992AdSpR..12d..47S,Bernstein1995ApJ...454..327B,Bernstein1999Sci...283.1135B, Strazzulla1997AdSpR..19.1077S, Palumbo1999A&A...342..551P} and/or quantum tunneling effects dominate \citep[e.g.][]{ meisner2016atom, andersson_tun, meisner_lambert_kast, molp_tun_abstractions, molp_tun_rivill, Molp_tun_nh2oh, Molp_tun_phosph, Molp_tun_addition}.

Among the different astrochemically relevant compounds, those called interstellar complex organic molecules (iCOMs) enjoy a particularly important position \citep{herbst_complex_2009,ceccarelli_seeds_2017}.
iCOMs are molecules with 6-13 atoms in which at least one is C and other heteroatoms (O, N, S...), rendering them the organic character. 
Examples of often detected iCOMs are acetaldehyde (CH$_3$CHO), formamide (NH$_2$CHO), methyl formate (CH$_3$OCHO) and dimethyl ether (CH$_3$OCH$_3$) \citep[e.g. see the latest reviews by][]{Jorgensen2020-ARA&A, Ceccarelli2022-PP7}.
Despite they are very small and simple molecules when compared to the terrestrial ones, iCOMs have gained great attention in the last two decades because some of them are actual biomolecule precursors, the possible molecular building blocks of biological systems, and are detected in star-forming regions which will eventually give rise to solar-like planetary systems \citep{Cazaux2003}. 
Therefore, iCOMs may represent primogenial organic chemistry and could have played a role in the emergence of life on Earth.
 
The first detections of iCOMs were obtained towards massive star-forming regions, in the so-called hot cores \cite[e.g.][]{Rubin1971-formamide,Blake1987-orion}, and only almost 30 years later in Solar-like protostar hot corinos \citep{Ceccarelli2000-glycine,Cazaux2003}.
More recently, iCOMs have been detected in cold prestellar cores \citep[e.g.][]{Bacmann2012, Cernicharo2012, Vastel2014-PSCiCOM, Scibelli2020-PSC}, protostellar outflows \citep[e.g.][]{Arce2008, Codella2009-B1ch3cn, Lefloch2017, Codella2017-B1formamide, simone_seeds_2020}, protoplanetary disks \citep{Oberg2015,Walsh2016-ch3oh,Favre2018,Ilee2021-iCOMdisks}, and even external galaxies \citep[e.g.][]{Henkel1987-CH3OHextragal,Muller2011-extragal,Martin2021-extragal}.
At present, more than 290 species have been detected in different astrophysical environments \citep[][and \url{https://cdms.astro.uni-koeln.de/classic/molecules}]{mcguire2021}. 

The formation of iCOMs has been postulated to occur in the gas phase and/or on the icy surfaces of grains. 
In the latter, a prevailing mechanism advocates the coupling between two radical species that were previously formed in situ due to UV incidence on the ice mantles \citep[e.g.][]{Garrod013-glycine,Iqbal2018,Suzuki2018}. 
However, it has recently been shown that radical-radical reaction on water ices may exhibit activation barriers depending on the ice composition, surface morphology, type of adsorption site, binding energy of the reactive radicals with the surface, and the orientation of the reactive radicals between them \citep[e.g.][]{enrique-romero_quantum_2022}. Furthermore, reactive radicals need to encounter via surface diffusion, which is affected by all these factors, and is not trivial due to the very low interstellar temperatures \citep[e.g.][]{enrique-romero_theoretical_2021}.  

Because of that, alternative grain-surface iCOM formation reaction mechanisms have been proposed such as the radical-ice one, in which a radical reacts directly with a component of the ice through an Eley-Rideal mechanism \citep{[e.g.][]rimola_can_2018, perrero_non-energetic_2022, Ferrero2023-C+CO}. 
In this context, a promising route towards iCOMs formation is based on the high reactivity of atomic C. 
On the surfaces of grains, the carbon atom, either in its ground electronic state (C, $^3$P) or in its cationic form (C$^+$, $^2$P), is indeed a highly reactive species, which reacts barrierlessly with several components.
Hence, the reaction of C atoms with molecules of the ice mantles could give rise to a variety of interstellar molecules, including iCOMs 
\citep{krasnokutski_low-temperature_2017, krasnokutski_condensation_2020, krasnokutski_pathway_2022, henning_experimental_2019, QASIM_CH4, molpeceres_carbon_2021, fedoseev_hydrogenation_2022,woon_quantum_2021,potapov_new_2021, Ferrero2023-C+CO}.

The formation of methane from the hydrogenation of carbon atoms is one of the first reactions evoked to occur on the icy surfaces of the interstellar grains \citep[e.g.][]{Tielens1982,cuppen_grain_2017}.
Only recently, the CH$_4$ formation on water ice by hydrogenation of C was observed experimentally when co-depositing atomic C and H with H$_2$O at cryogenic temperatures \citep{QASIM_CH4}. 
Similarly, the formation of CH$_4$ was also demonstrated experimentally by the reaction of C with H$_2$ on water ice \citep{Lamberts2022_methane}. 
Moreover, in a combined theoretical and experimental work, it was shown that C ($^3$P) adsorption on water ice mantles reacts with the oxygen atom of an icy water molecule forming a C-OH$_2$ species, stable in the electronic triplet excited state, which upon a water-assisted proton transfer followed by an intersystem crossing (ISC) towards the fundamental singlet state evolves into formaldehyde (H$_2$CO) \citep{molpeceres_carbon_2021}.
Finally, a very recent experimental paper evidenced possible diffusion of carbon atoms on the surface of interstellar ice \citet{tsuge2023surface}.
These works, thus, show how complicated carbon chemistry on interstellar ice is, and that different outcomes, depending on a variety of factors, are possible upon adsorption.

The above-mentioned theoretical studies show the outermost importance of performing accurate quantum mechanical (QM) computations to correctly understand the experimental results (which will be done here in Section \ref{sec:discussion}). An analysis of the electronic structure of the $^3$C-OH$_2$ species indicates that the unpaired electrons (responsible for the triplet electronic state) remain mostly localised on the C end. 
Thus, the enhanced reactivity of the initial C ($^3$P) atom is transferred to the newly formed $^3$C-OH$_2$ species. 
This is of relevance because it opens the possibility that $^3$C-OH$_2$ is an activated species that can trigger the formation of interstellar molecules through non-energetic pathways, presenting likely very small or no barriers. 

In this work, we present results based on QM simulations to verify this hypothesis. 
This has been done by studying, first, the adsorption of atomic C on the surfaces of interstellar water icy grains, in which the $^3$C-OH$_2$ species forms, followed by its reactivity with relatively abundant interstellar species of different nature, that is: 
i) closed-shell species, abundant on the water-dominated ices (NH$_3$, CO, CO$_2$, and H$_2$); 
ii) atoms that can be found as adsorbed species on the ice surfaces (H, N, and O); 
and iii) molecular radicals also trapped on the ices (CH$_3$, NH$_2$, and OH), which could be the product of the partial atoms/molecules hydrogenation \citep[e.g.][]{Taquet2012-grainoble} or the in situ formation by energetic (UV photons or cosmic-rays particles) irradiation \citep[e.g.][]{Garrod013-glycine}.

The article is organised as follows.
Section \ref{sec:comp-details} describes the adopted methodology, Section \ref{sec:results} the obtained results, Section \ref{sec:discussion} discusses the consequences of these results and Section \ref{sec:conclusions} concludes the article.

\section{Computational details}\label{sec:comp-details}

All the simulations are grounded on the density functional theory (DFT) and have been performed using the Orca software \citep{neese_orca_2020}. The range-separated DFT hybrid $\omega$B97X-V method has been used due to its overall good performance among other hybrid functionals \citep{goerigk_look_2017}. Dispersion interactions are taken into account by employing the VV10 nonlocal correlation \citep{vydrov_nonlocal_2010}. Ahlrich’s def2-TZVP has been used as the basis set for all the calculations \citep{weigend_balanced_2005}. It is worth mentioning that similar functionals (i.e., $\omega$B97X-D3 and $\omega$B97X-D4, only differing in the definition for the description of dispersion interactions) have been shown to provide results in good agreement with the highly correlated CCSD(T) results (giving errors in energy barriers less than 10\%), and hence being accurate enough to describe similar energy barrier reactions as those presented in this work \citep{perrero_non-energetic_2022, Ferrero2023-C+CO}. Geometry optimisations have been carried out employing the geometrical counterpoise correction (GCP) \citep{kruse_geometrical_2012}, to palliate the basis set superposition error (BSSE) \citep{liu_accurate_1973}. Open shell systems have been treated with the unrestricted formalism. Electron spin densities and net atomic charges on the atoms have been obtained through a Löwdin population analysis.

\begin{figure}
    \centering
    \includegraphics[width=\linewidth]{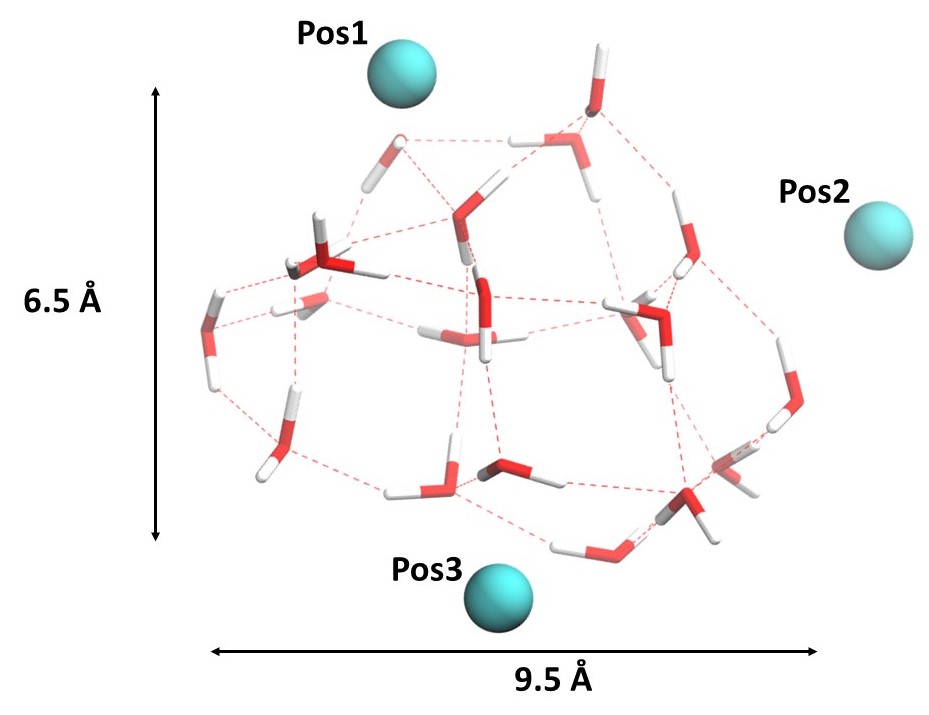}
    \caption{The (H$_2$O)$_{20}$ cluster model used in this work, optimised at $\omega$B97X-V/def2-TZVP, which also shows the size dimensions. The different positions in which the C atom has been adsorbed (Pos1, Pos2, and Pos3) are also shown.}
    \label{fig:cluster_pos}
\end{figure}

The interstellar ice has been modelled by adopting a cluster approach. It consists of a cluster model of 20 water molecules taken from \citet{shimonishi_adsorption_2018} and optimised at our level of theory. The cluster has an ellipsoid-like shape, with axes of approximately 6.5 and 9.5 {\AA} (see Fig. \ref{fig:cluster_pos}). Considering that the adsorption of atomic C and the subsequent reactivity are local surface phenomena (i.e., occurring on single binding and reaction sites), this cluster has been chosen as a good trade-off between the representativity of an actual interstellar amorphous water ice surface and the computational cost of the calculations.

Harmonic vibrational frequencies have been calculated to i) characterise the nature of the stationary points of the calculated potential energy surfaces (PES), i.e., reactants, products, and intermediates as minima of the PES, and transition states as first-order saddle points of the PES, and ii) take into account the vibrational zero-point energy (ZPE) of each stationary point to obtain ZPE-corrected energetics. 
The ZPE-corrected interaction energy of atomic C on the water ice model was calculated as:

\begin{equation}
    \Delta \text{H}(0)= E_{complex} - E_{ice} - E_{C}
\end{equation}

where $E_{complex}$ and $E_{ice}$ are the ZPE-corrected absolute potential energies for the carbon adsorbed on the water ice cluster and for the isolated optimised water ice cluster, and $E_{C}$ is the absolute potential energy for the isolated carbon atom. 
Transition state structures have been localized adopting the climbing image nudged elastic band (CI-NEB) technique implemented in Orca \citep{asgeirsson_nudged_2021}. ZPE-corrected energy barriers of the explored reactions have been calculated as:

\begin{equation}
    \Delta \text{H}^\ddag(0) = E_{TS} - E_{min}
\end{equation}

where $E_{TS}$ and $E_{min}$ are the ZPE-corrected absolute potential energies for the transition state and the minimum structure of the reaction, respectively. Note that at 0 K, the absolute ZPE-corrected energy is equal to the absolute enthalpy, i.e., E = H(0), and hence our notation in terms of enthalpy variation at 0 K for the interaction energies and energy barriers.
The visual molecular dynamics (VMD) software was used for the display and manipulation of the structures and for image creation \citep{humphrey_vmd_1996}.

\section{Results} \label{sec:results}

\subsection{Carbon condensation on water ice} \label{subsec:C-condensation}

\begin{figure}
    \centering
    \includegraphics[width=\linewidth]{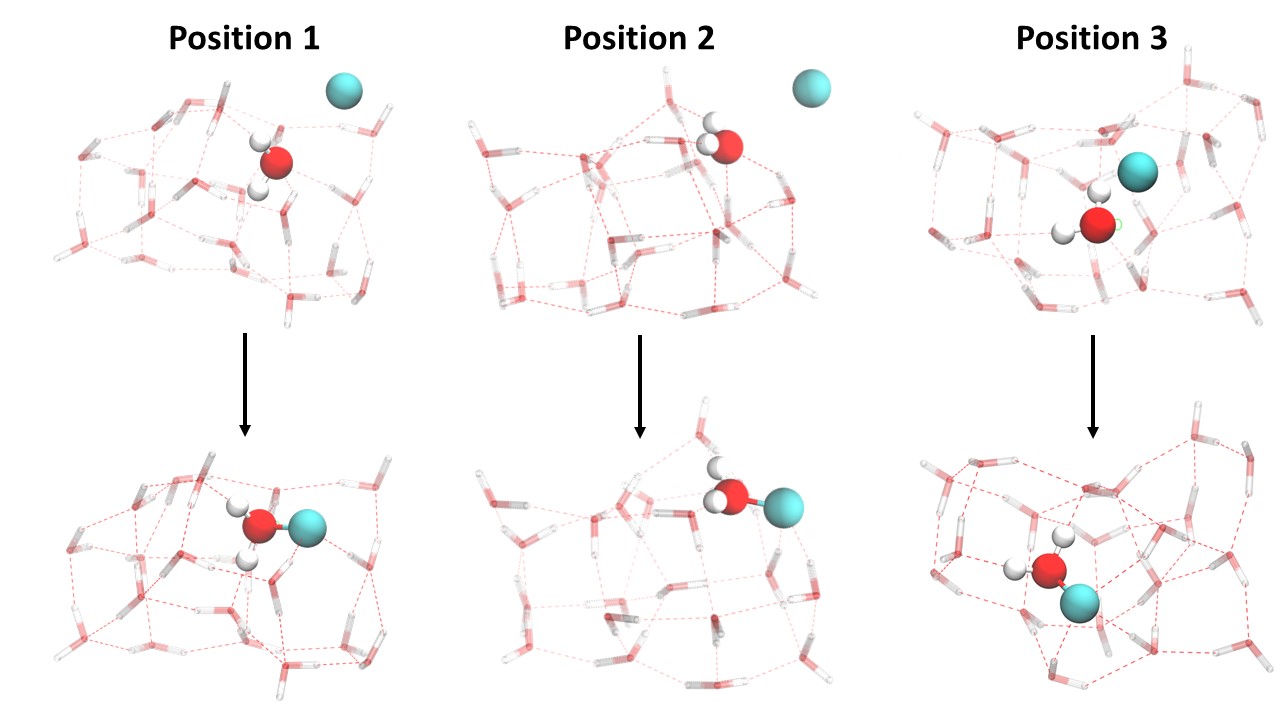}
    \caption{$\omega$B97X-V-optimised structures for C atom on the (H$_2$O)$_{20}$ cluster model, which results in the formation of a C-OH$_2$ species. 
    The initial guess structures of the three different positions (Pos1, Pos2, and Pos3) are also shown. 
    The reactive species are represented in ball and stick, while the rest of the water molecules of the cluster are in tubes. The color code for the atoms is H in white, C in pale green and O in red.}
    \label{fig:c_ads}
\end{figure}

To simulate the adsorption of atomic C ($^3$P) on the water ice, the two partners (i.e., the C atom and the (H$_2$O)$_{20}$ cluster) have been placed 3.0 {\AA} apart from each other, and the geometries of which were then optimised. 
Three different initial guess structures, differing in the position of the C atom around the cluster, have been used (see Fig. \ref{fig:cluster_pos}). The selection of the positions was based on considerations of surface morphology and coordination (according to the Fowler rules) of the H$_2$O molecule that reacts with atomic C. That is: in Pos1, C reacts with an O-undercoordinated water molecule belonging to a small cavity of the surface, in Pos 2 with a completely coordinated water molecule, and in Pos 3 with an O-undercoordinated water molecule but belonging to a flat part of the surface.
As a result, for each initial position, the spontaneous reaction between the C atom and a water molecule of the ice takes place during the geometry optimisations, forming the C-OH$_2$ species (depicted in Fig. \ref{fig:c_ads}), which due to the spin conservation rule is in the triplet electronic state.

The formation of this $^3$C-OH$_2$ species was already observed in the computational studies of \citet{HWANG1999143}, \citet{shimonishi_adsorption_2018}, \citet{molpeceres_carbon_2021} and \citet{Lamberts2022_methane}. 
For the three initial positions considered in this work, the calculated interaction energies ($\Delta$H(0)) of the C atom with the water cluster are reported in Tab. \ref{tab:deltaH}, alongside the C-O bond lengths, charges and spin densities on the C and O atoms of the C-OH$_2$ species.

\begin{table}[ht]
\centering
    \caption{Computed ZPE-corrected interaction energies ($\Delta$H(0), in kJ mol$^{-1}$), of the C atom with the water ice surface cluster model, considering the three positions (Pos1, Pos2, and Pos3). 
    The C-O bond lengths (in {\AA}), the Löwdin net atomic charges and the electronic spin densities of the C and O atoms of the optimised $^3$C-OH$_2$ species on the water cluster are also listed.}
\label{tab:deltaH}
\begin{tabular}{ccccccc}
\hline
Position & $\Delta$H(0) & C-O length & Charges & \multicolumn{3}{c}{Spin densities} \\ 
         &              &            &         & O    & C    & O \\ \hline 
Pos1     & -116.2       & 1.49       & -0.52   & 0.06 & 1.77 & 0.08 \\
Pos2     & -115.0       & 1.49       & -0.53   & 0.09 & 1.73 & 0.10  \\
Pos3     & -110.2       & 1.50       & -0.55   & 0.08 & 1.72 & 0.09 \\ 
\hline
\end{tabular}
\end{table}

Both the calculated $\Delta$H(0) and the C-O distances are in very good agreement with those reported by \citet{shimonishi_adsorption_2018} and \citet{molpeceres_carbon_2021}. 
The values reported by \citet{Lamberts2022_methane} are lower than the ones computed here. 
This is because the systems adopted by the authors refer to the coupling of the C atom with one isolated molecule (-52.9 kJ mol$^{-1}$) and with the (H$_2$O)$_3$ water trimer cluster (-95.2 kJ mol$^{-1}$). 
Interestingly, by increasing the cluster size, the formation energy of the reactive carbon centre increases considerably, tending to our obtained values. 
Because of the coupling reaction of C ($^3$P) with a water molecule of the ice gives rise to the $^3$C-OH$_2$, this adsorption event can be classified as chemisorption. 
According to the electronic spin densities, the C atom retains its two unpaired electrons upon the $^3$C-OH$_2$ formation and, accordingly, this species is expected to be reactive through the C atom. 
Thus, the $^3$C-OH$_2$ species can be understood as a carbon reactive centre, that is, like an activated complex that reacts through its C atom with other species that can diffuse in its proximity or land from the gas phase on surface sites close to it. 
Moreover, according to the Löwdin atomic charges, the carbon atom presents a negative net charge, which renders this atom a good H-bond acceptor, like the oxygen in water. 
These two facts will have important consequences on the chemical properties of the carbon reactive centre, as we will see in the following.

\subsection{Reactivity of the carbon reactive centre with abundant closed-shell molecules: NH$_3$, CO, CO$_2$ and H$_2$} \label{subsec:CreacClosedShellMol}

\begin{figure}
    \centering
    \includegraphics[width=\linewidth]{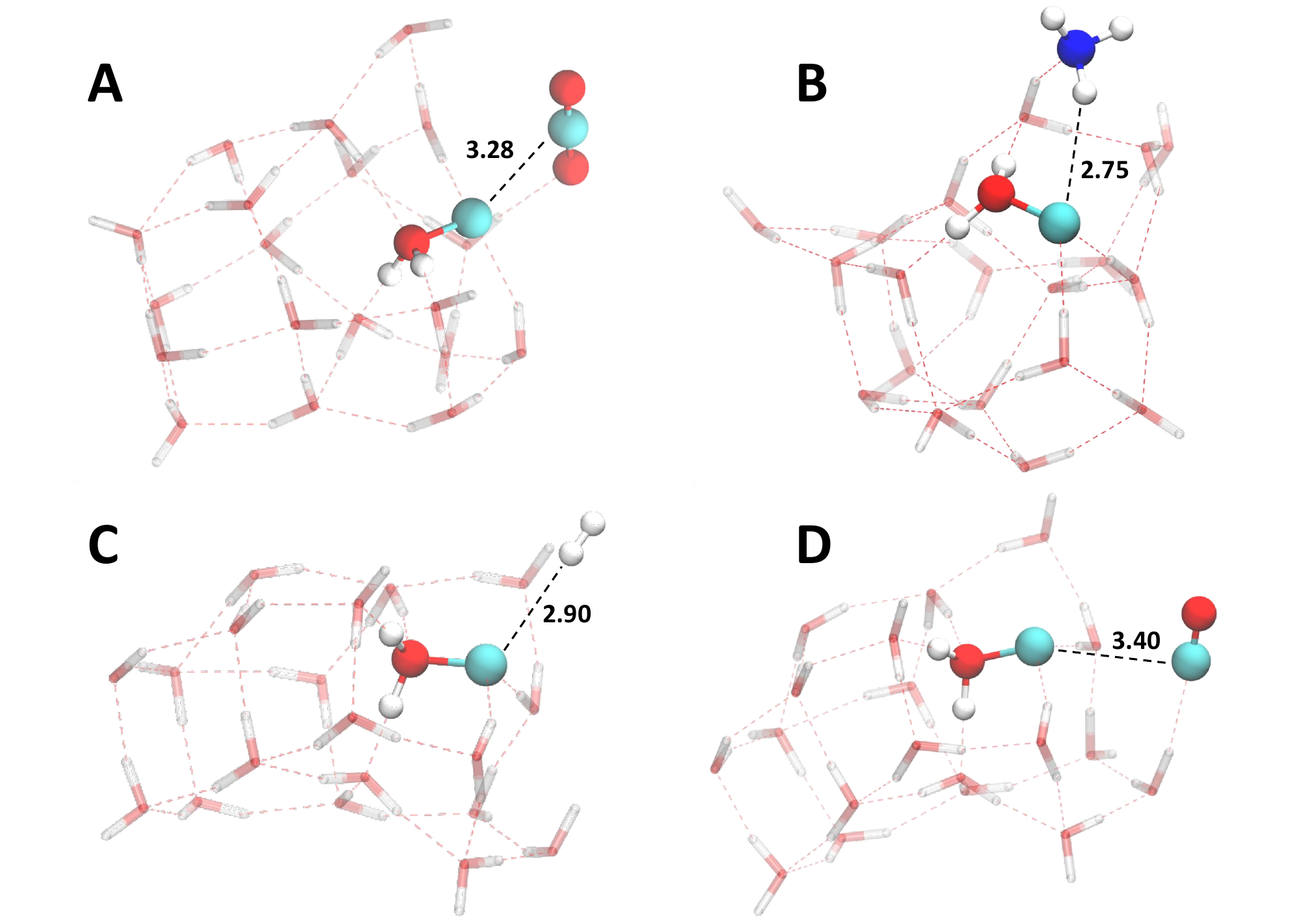}
    \caption{Optimised geometry of the closed shell species on the water cluster with the carbon reactive centre. Panel A) CO$_2$ optimised geometry in Pos1. Panel B) NH$_3$ optimised geometry in Pos2. Panel C) H$_2$ optimised geometry in Pos3. Panel D) CO optimised geometry in Pos1. In all the panels the adsorbed molecules plus the C-OH$_2$ species are highlighted in ball and stick representation. Distances (in {\AA}) between the carbon atom of the C-OH$_2$ moiety and the closest atom of the adsorbed species are highlighted with a black dashed line. The color code for the atoms is H in white, C in pale green, N in blue and O in red.}
    \label{fig:closed_shell}
\end{figure}

The reactivity with the relatively abundant molecular constituents of the interstellar icy mantles (NH$_3$, CO, CO$_2$) and the most abundant molecule of the ISM (H$_2$) has been studied by placing them 2.5 {\AA} far from the C atom of the $^3$C-OH$_2$ species, considering the three different positions as depicted in Fig. \ref{fig:c_ads}. 
With any of the molecules tested, no spontaneous reaction is observed during the geometry optimisation (performed at a total triplet electronic state), resulting in the systems represented in Fig. \ref{fig:closed_shell}. 
Results indicate that the molecular species prefer to interact through hydrogen bonds (for the CO, CO$_2$ and NH$_3$ cases) or dispersion interactions (for the H$_2$ case) with either the water molecules of the ice or the C atom of $^3$C-OH$_2$ (for the NH$_3$ case) due to its negative atomic charge (see above), in this latter situation C acting similarly to a dangling oxygen of water ice. 

\subsection{Reactivity of the carbon reactive centre with abundant atoms: H, N and O} \label{subsec:CreacAtoms}

Based on the interstellar abundances and their mobility on the ice mantles, the highest probability that the $^3$C-OH$_2$ has a reactive encounter is with atomic hydrogen, followed by atomic nitrogen and oxygen. 
Therefore, the reactions between the carbon reactive centre with these three atomic species have been studied.

\begin{figure}
    \centering
    \includegraphics[width=\linewidth]{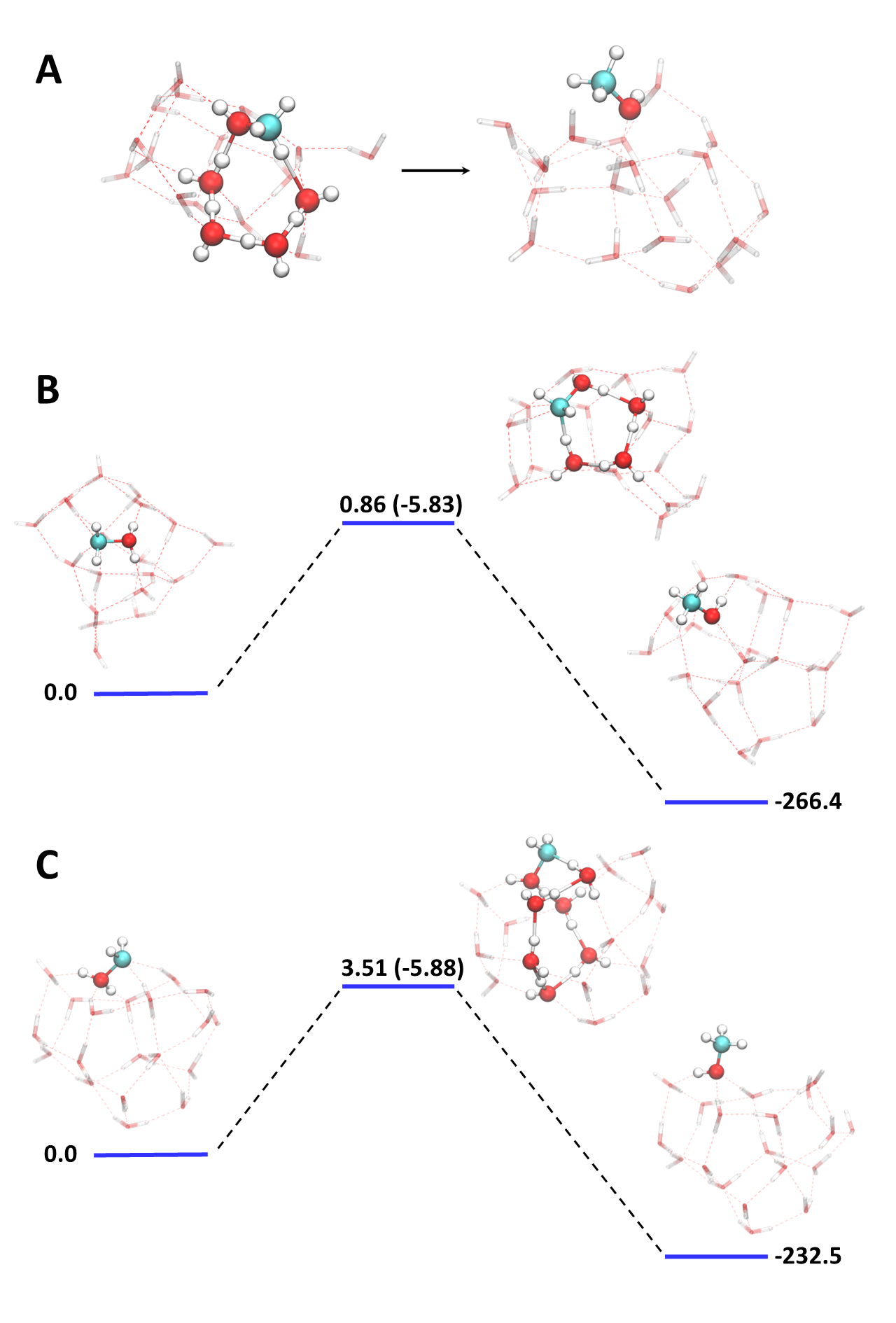}
    \caption{Panel A) Snapshot of the geometry optimisation representing the water-assisted proton transfer mechanism for the H$_2$C-OH$_2$ $\rightarrow$ CH$_3$-OH reaction and of the final product of Pos1. 
    Panel B) and C) Potential energy surfaces (in kJ mol$^{-1}$) for the water-assisted proton transfer of Pos2 and Pos3. 
    At the TS structure, the $\Delta$H(0) is displayed in parenthesis indicating barrierless reactions. 
    In every panel, the H$_2$C-OH$_2$ species, the water molecules involved in the proton transfer as well as methanol are highlighted in ball and stick representation. The color code for the atoms is H in white, C in pale green and O in red.}
    \label{fig:shuttles}
\end{figure}

\medskip
\noindent
\textit{H:}
To simulate the reaction with H, an H atom was placed at 2.5 {\AA} apart from the $^3$C-OH$_2$ derived from Pos1, Pos2 and Pos3, and the geometries of these initial guess structures were optimised in a doublet spin state, as it is the reactive multiplicity (in contrast to the unreactive quartet state).
In this process, the H atom spontaneously makes a chemical bond with the carbon atom of $^3$C-OH$_2$, forming the $^2$HC-OH$_2$ radical species, which remains adsorbed on the ice cluster. 
This species, in a similar way as the $^3$C-OH$_2$ one, can also be regarded as a carbon reactive centre due to its doublet radical character, which is still mainly localised on the carbon end. 
Therefore, a second H atom addition was also performed. 
As before, this second H atom forms spontaneously a chemical bond with the carbon reactive centre, forming a closed shell H$_2$C-OH$_2$ species. Interestingly, the optimisation process of the second H addition for the system arising from Pos1 is associated with a spontaneous water-assisted proton transfer, in which one proton linked to the water-end of the H$_2$C-OH$_2$ species is transferred to the C atom through water molecules of the ice model, forming finally methanol in a barrierless way. 
That is, H$_2$C-OHH* $\rightarrow$ CH$_2$H*-OH, in which the H* is the transferred atom. 
Figure \ref{fig:shuttles}A shows a snapshot of the geometry optimisation, in which the water-assisted proton transfer process is highlighted. 

In view of the possibility of this conversion to form methanol, we run NEB calculations for the other two cases (Pos2 and Pos3) to characterise the minimum energy path and calculate the energy barrier for the formation of methanol adopting a similar mechanism. 
The obtained results (shown in Fig. \ref{fig:shuttles}B and C) indicate that, in both cases, the process presents a very small potential barrier ($\leq$ 4 kJ mol$^{-1}$), which moreover becomes submerged when the ZPE-corrections are introduced.

Thus, in practice, the two surface reactions studied here are (the multiplicity of each species is marked in the left super index):

%

\vspace{0.1in}

\ce{^2\text{H} \ch{+} ^3\text{C-OH}_2 \ch{->} ^2\text{HC-OH}_2} 

\ce{ ^2\text{H} \ch{+} ^2\text{HC-OH}_2 \ch{->} \text{CH}_3\text{OH}} 

\vspace{0.1in}

These results indicate, therefore, that even at the cryogenic temperatures of the cold molecular clouds, the formation of CH$_3$OH through this chemical route is efficient due to the high reactivity of the carbon reactive centre alongside the abundance and high mobility of the hydrogen atom on the interstellar ice surfaces. 

\begin{figure}
    \centering
    \includegraphics[width=\linewidth]{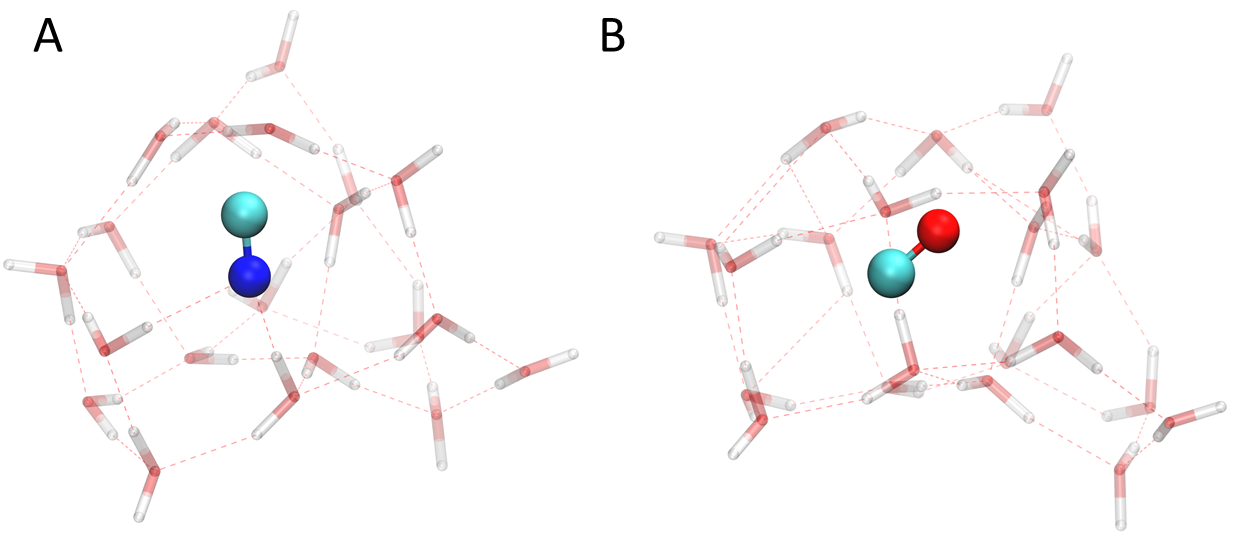}
    \caption{Final products for the reaction of atomic N (panel A) and O (panel B) with the $^3$C-OH$_2$ carbon reactive centre on water ice surfaces, which result in the formation of CN$^{\cdot}$ and CO, respectively. These final species are highlighted in a ball and stick representation. The color code for the atoms is H in white, C in pale green, N in blue and O in red.}
    \label{fig:COCN}
\end{figure}

\medskip
\noindent
\textit{N and O:}
Now, let us focus on the reactions with atomic N and O, which have been studied in the same way as H. 
The fundamental electronic state of atomic nitrogen is $^4$S and of oxygen $^3$P. 
Thus, the total reactive electronic spin multiplicities with $^3$C-OH$_2$ are doublet and singlet, respectively. 
During the optimisation process, the atoms approach the C-OH$_2$ and, in a spontaneous and concerted way, they establish a chemical bond with the carbon atom, forming the radical CN$^{\cdot}$ and the neutral CO species, in both cases breaking the original C-O bond of the carbon reactive centre (see Fig. \ref{fig:COCN}) and releasing H$_2$O. 
In summary, the two surface reactions studied here are:


\vspace{0.1 in}

\ce{\text{N}(^4\text{S}) \ch{+} ^3\text{C-OH}_{2} \ch{->} ^2\text{CN}^{$\cdot$} \ch{+} \text{H}_2\text{O}}

\ce{\text{O}(^3\text{P}) \ch{+} ^3\text{C-OH}_{2} \ch{->} ^1\text{CO} \ch{+} \text{H}_2\text{O}}  

\vspace{0.1 in}

The reason why in these cases the OH$_2$ moiety of the carbon reactive centre detaches from the C atom is of thermodynamic nature: there is a large energy gain by forming the triple \ch{C+O} and \ch{C+N} chemical bonds (reaction energies of -684 and -958 kJ mol$^{-1}$, respectively), in detriment to breaking the single C-O bond. Such reaction energies are expected to be efficiently absorbed and dissipated throughout the ice (as shown in computational works regarding energy dissipation of large exothermic reactions (e.g., \citep{pantaleone_chemical_2020, pantaleone_h2_2021,Molpeceres2023}) due to vibrational coupling \citep{FerreroNH3-2023}. However, the amount of energy released by the reactions is very large, and accordingly, side phenomena cannot be discarded. For the case of CO, chemical desorption (namely, the desorption of a chemical species upon its formation owing to the local heating caused by the exothermicity of the reaction) could occur, since the binding energy of CO on water is relatively low \citep{ferrero_binding_2020}. For the case of CN$^{\cdot}$, the reaction energy could be used to proceed with subsequent chemical reactions. Indeed, in \cite{rimola_can_2018}, it was postulated the reactivity of radical CN$^{\cdot}$ with icy water molecules, forming first NH$_2$CO and followed by an H addition to finally form NH$_2$CHO. This reaction could well take place once CN$^{\cdot}$ forms by the reaction of N with C-OH$_2$.

\subsection{Reactivity of the carbon reactive centre with abundant radicals: OH, CH$_3$ and NH$_2$} \label{subsec:CreacRadicals}

The reactions of these three radical species have been studied adopting the same approach as that used for the atomic species, that is, by placing the radicals 2.5 {\AA} apart from the $^3$C-OH$_2$ and optimising the system in the total reactive electronic state, in this case, the doublet one. 

\medskip
\noindent
\textit{OH and CH$_3$:}
For the addition of OH and CH$_3$ to $^3$C-OH$_2$, we found the spontaneous formation of a new chemical bond between the carbon atom of $^3$C-OH$_2$ and the O and C atoms of OH and CH$_3$, respectively, during the geometry optimisation, in which the $^2$HO-C-OH$_2$ and $^2$H$_3$C-C-OH$_2$ species are formed. 
We observed this behaviour in all cases regardless of the initial geometry.

The barrierless reaction paths observed are associated with the following chemical reactions:


\vspace{0.1 in}
\ce{^2\text{OH} \ch{+} ^3\text{C-OH}_2 \ch{->} ^2\text{HO-C-OH}_2} 

\ce{^2\text{CH}_3 \ch{+} ^3\text{C-OH}_2 \ch{->} ^2\text{H}_3\text{C-C-OH}_2}
\vspace{0.1 in}

These reactions are very similar to the first H addition to the carbon reactive centre and, thus, the unpaired electron in $^2$HO-C-OH$_2$ and $^2$H$_3$C-C-OH$_2$ is mostly localised on the carbon atom of the C-OH$_2$ moiety. 
Because of this similarity, a second reaction of the $^2$HO-C-OH$_2$ and $^2$H$_3$C-C-OH$_2$ with an H atom has been investigated to check additional chemical reactivity, which in turn can evolve to other more stable species through a non-energetic proton shuttle mechanism.

\begin{figure}
    \centering
    \includegraphics[width=\linewidth]{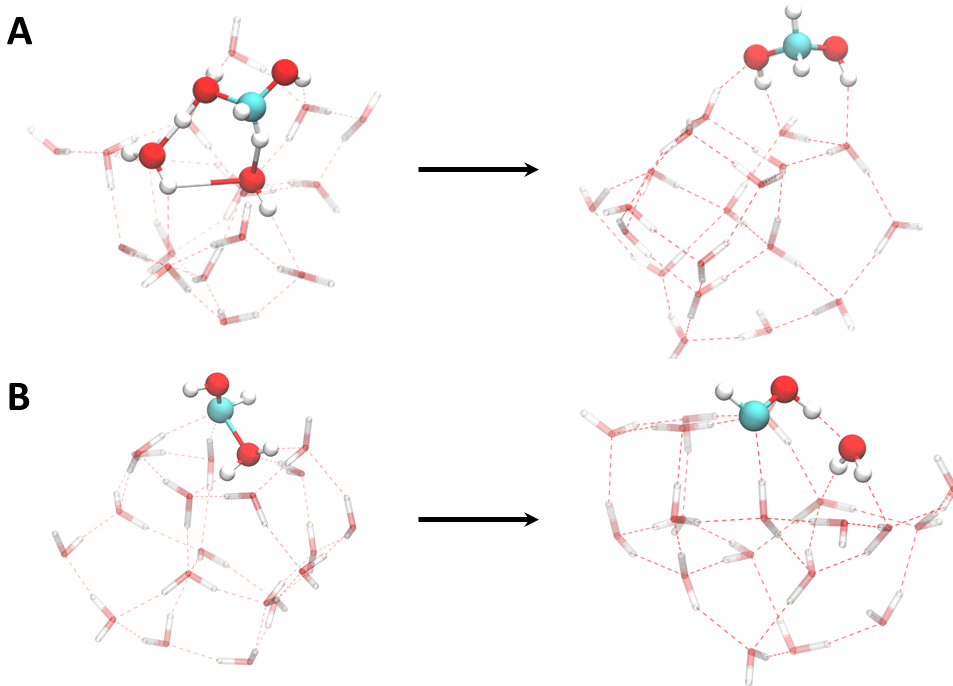}
    \caption{Snapshots of the reactions of the OH radical with the $^3$C-OH$_2$ carbon reactive centre on water ice surfaces. In panel A, the geometry optimisation representing the water-assisted proton transfer mechanism for the HO-H$_2$C-OH$_2$ $\rightarrow$ HO-CH$_2$-OH reaction and the final product are represented. The HO-H$_2$C-OH$_2$ species and the water molecules involved in the proton transfer are highlighted in ball and stick representation (left structure) as well as methandiol in the optimised position (right structure). In panel B, snapshots of the HO-HC-OH$_2$ intermediate (left panel) during the C-O bond breaking and of the HCOH product (right panel) in its optimised geometry. The two species are highlighted in ball and stick representation. The color code for the atoms is H in white, C in pale green and O in red.}
    \label{fig:ch3_attack}
\end{figure}

For the H addition to $^2$HO-C-OH$_2$, two different products have been found. 
From Pos1 and Pos2, methanediol (OH-CH$_2$-OH) compound is formed by a barrierless water-assisted proton transfer (see Fig. \ref{fig:ch3_attack}A). 
From the remaining position (Pos3), the hydroxycarbene (HOCH) species is formed after the cleavage of the original C-O bond belonging to the $^3$C-OH$_2$ species (see Fig. \ref{fig:ch3_attack}B). 
Both processes were observed during the geometry optimisation and, thus, they can be considered spontaneous. 
The observed chemical reactions are: 


\vspace{0.1in}

\ce{^2\text{HO-C-OH}_2 \ch{+} ^2\text{H} \ch{->} ^1\text{HO-CH}_2\text{-OH}}

\ce{ ^2\text{HO-C-OH}_2 \ch{+} ^2\text{H} \ch{->} ^1\text{HCOH} \ch{+} \text{H}_2\text{O}}

\vspace{0.1in}

\begin{figure}
    \centering
    \includegraphics[width=\linewidth]{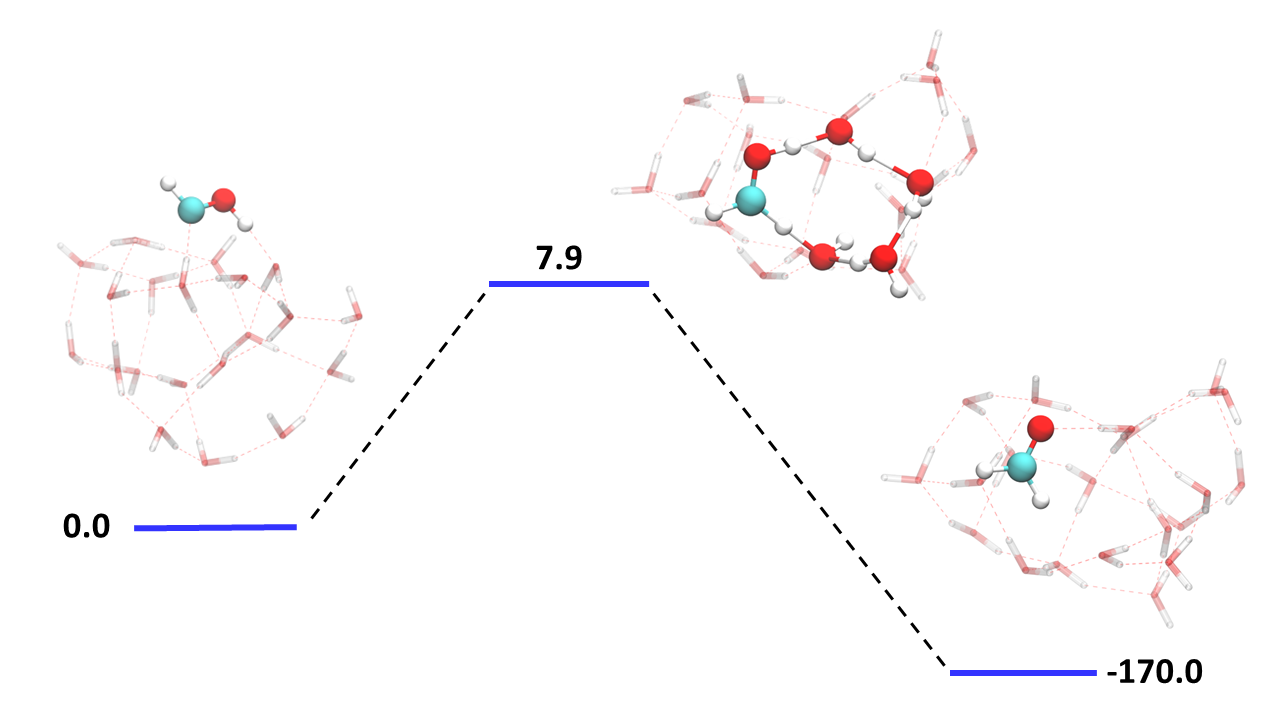}
    \caption{Energy diagram (e.g. reactant, TS and product) representing the water-assisted proton transfer mechanism for the HCOH $\rightarrow$ H$_2$CO reaction. ZPE-corrected energies are reported in kJ mol$^{-1}$. The HCOH and H$_2$CO species in their optimised positions as well as the water molecules involved in the proton transfer are highlighted in ball and stick representation. The color code for the atoms is H in white, C in pale green and O in red.}
    \label{fig:h2co_shuttle}
\end{figure}

Interestingly, the formation of HCOH (Fig. \ref{fig:ch3_attack}B) was highlighted in the work of \citet{molpeceres_carbon_2021} to be a precursor species of formaldehyde (i.e., HCOH $\rightarrow$ H$_2$CO). For this reaction, they found a variety of energy barriers ranging (including ZPE correction) from submerged barriers up to 12 kJ mol$^{-1}$, depending on the binding site of the C atom. 
In this work, to obtain the activation energy of the HCOH $\rightarrow$ H$_2$CO conversion, a NEB calculation was performed, obtaining a ZPE-corrected energy barrier of 7.9 kJ mol$^{-1}$ (see Fig. \ref{fig:h2co_shuttle}), which lies within the range of values found in \citet{molpeceres_carbon_2021}.
Therefore, this water-assisted proton transfer is not barrierless, in contrast to what we found in the other cases. 
Nevertheless, it is important to notice that the estimated barrier is a classical potential barrier and that nuclear quantum effects (not accounted for in this work) can largely affect the kinetics of this process.

To have an estimate upon this point, we have computed the crossover temperature (namely, the temperature below which tunnelling dominates) adopting the formulation of \citet{fermann2000modeling}. 
Our estimation gives a crossover temperature of 194 K and therefore, this water-assisted proton transfer cannot be ruled out and be feasible at the cryogenic temperature of the ISM via collective proton tunnelling mechanisms \citep{drechsel-grau_quantum_2014}.

For the hydrogenation of $^2$H$_3$C-C-OH$_2$ (the resulting product by the reaction of CH$_3$ with the carbon reactive centre), we found that only in Pos3 the H addition proceeds in a barrierless way to form ethanol, whereas no spontaneous reactive events were observed in Pos1 and Pos2. This highlights the strong dependence of this kind of processes on the different carbon binding sites (e.g., surface morphology, binding energies, or binding environments). 

The reaction forming ethanol can be written as:


\vspace{0.1in}

\ce{^2\text{H}_3\text{CC-OH}_2 \ch{+} ^2\text{H} \ch{->} \text{H}_3\text{CCH}_2\text{OH}}

\vspace{0.1in}

\begin{figure}
    \centering
    \includegraphics[width=\linewidth]{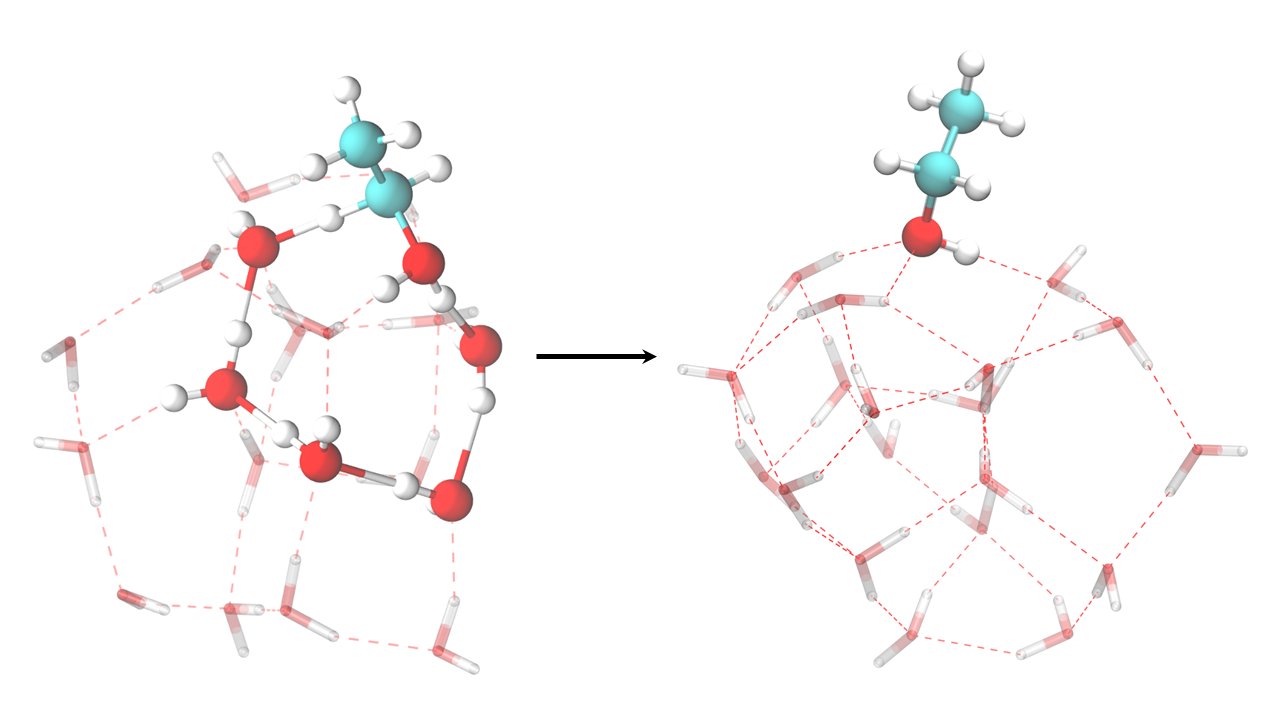}
    \caption{Snapshot of the geometry optimisation representing the water-assisted proton transfer mechanism for the CH$_3$-H$_2$C-OH$_2$ $\rightarrow$ CH$_3$-CH$_2$-OH reaction and of the final product. The CH$_3$-CH$_2$-OH$_2$ species and the water molecules involved in the proton transfer are highlighted in ball and stick representation (left structure) as well as ethanol in the optimised position (right structure). The color code for the atoms is H in white, C in pale green and O in red.}
    \label{fig:ch3_shuttle}
\end{figure}

Like several cases shown in this work, the H addition triggers a spontaneous water-assisted proton shuttle that allows ethanol formation as the final product, as shown in Fig. \ref{fig:ch3_shuttle}.

\begin{figure}
    \centering
    \includegraphics[width=\linewidth]{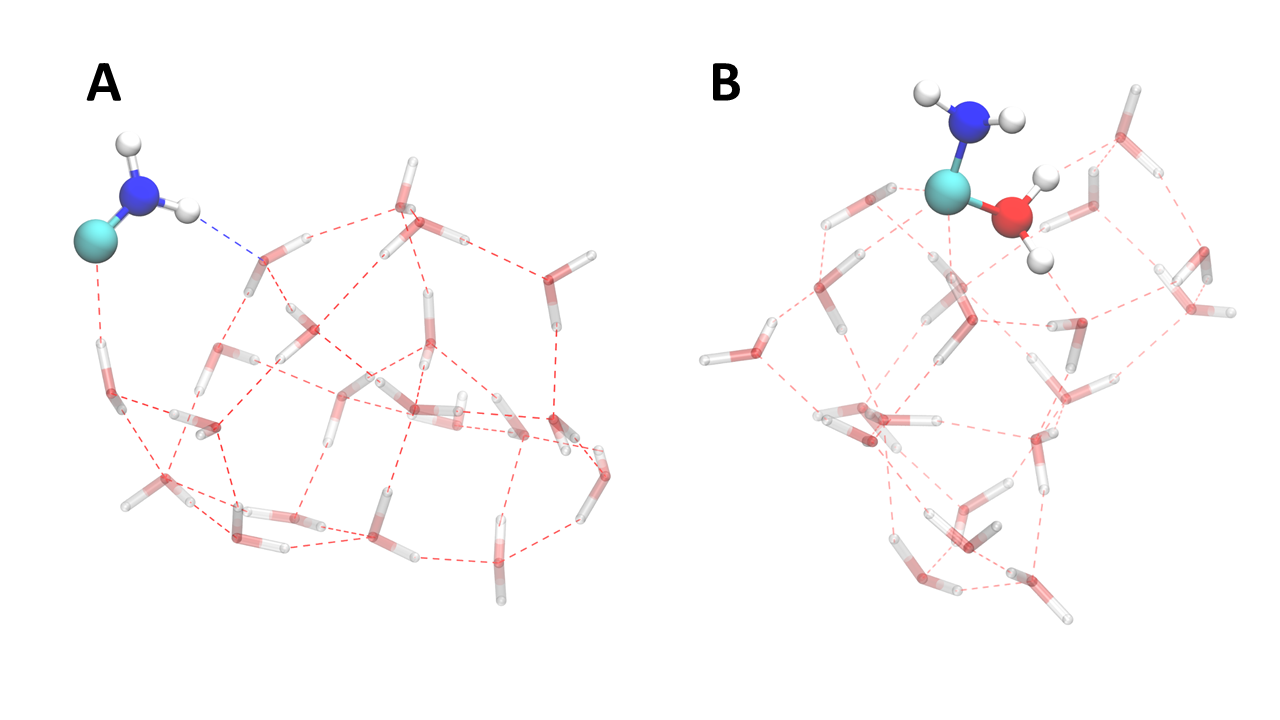}
    \caption{Panel A: $^2$C-NH$_2$ species highlighted in ball and stick representation in its optimised geometry. Panel B) $^2$H$_2$N-COH$_2$ species highlighted in ball and stick representation in its optimised geometry. The color code for the atoms is H in white, C in pale green, N in blue and O in red.}
    \label{fig:nh2_species}
\end{figure}

\medskip
\noindent
\textit{NH$_2$:}
Finally, as far as the NH$_2$ radical is concerned, its addition to $^3$C-OH$_2$ behaves differently than those for OH and CH$_3$. 
Indeed, the addition of NH$_2$ to Pos1 and Pos2 cleaves the original C-O bond during the optimisation process and leads to the formation of the $^2$C-NH$_2$ species (see Fig. \ref{fig:nh2_species}A). 
The reason is that, like in the O and N additions, the formed C-N bond presents an enhanced stability (reaction energy of -473 kJ mol$^{-1}$) with respect to the single C-O bond of the carbon reactive centre and, accordingly, in the energy balance, the process is thermodynamically favourable. 
That is:


\vspace{0.1in}
\ce{^2\text{NH}_2 \ch{+} ^3\text{C-OH}_2 \ch{->} ^2\text{NH}_2\text{C} \ch{+} \text{H}_2\text{O}}
\vspace{0.1in}

From Pos3, however, the NH$_2$ radical adds to the carbon reactive centre in a barrierless way and without breaking the original C-O bond, hence forming the $^2$H$_2$N-COH$_2$ species (see Fig. \ref{fig:nh2_species}B). 
The reaction is, thus:


\vspace{0.1in}
\ce{^2\text{NH}_2 \ch{+} ^3\text{C-OH}_2 \ch{->} ^2\text{NH}_2\text{-C-OH}_2}
\vspace{0.1in}

For both species formed (namely, $^2$NH$_2$C and $^2$NH$_2$-C-OH$_2$), we studied the addition of one H atom. 
Irrespective of the reactants, in both cases, the hydrogenation leads to the spontaneous formation of NH$_2$CH. 
On $^2$NH$_2$C, the H atom adds directly to the C atom, while on $^2$NH$_2$-C-OH$_2$, the H addition to the C atom induces the cleavage of the C-O bond, releasing the initial icy H$_2$O molecule. 

\begin{figure}
    \centering
    \includegraphics[width=\linewidth]{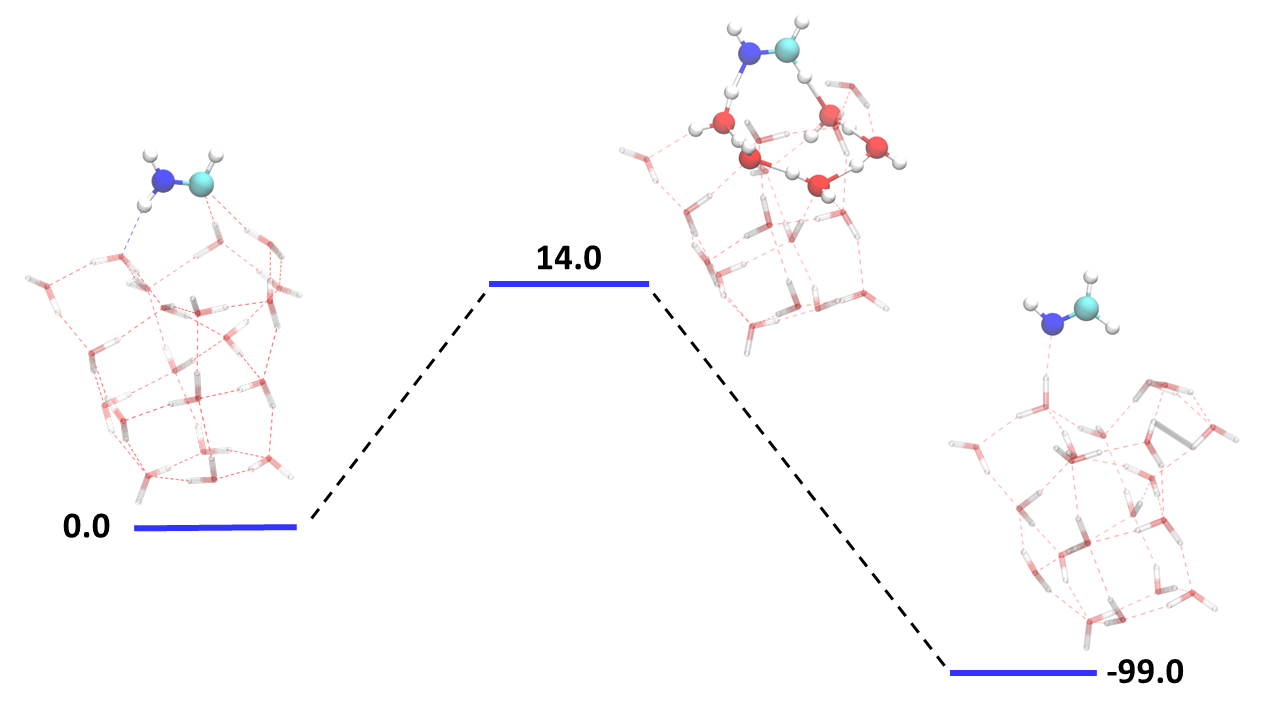}
    \caption{Energy diagram (e.g reactant, TS and product) representing the water-assisted proton transfer mechanism for the HCNH$_2$ $\rightarrow$ H$_2$CNH reaction. ZPE-corrected energies are reported in kJ mol$^{-1}$. The HCNH$_2$ and methanolimine (H$_2$CNH) species in their optimised positions, as well as the water molecules involved in the proton transfer, are highlighted in ball and stick representation. The color code for the atoms is H in white, C in pale green, N in blue and O in red.}
    \label{fig:nh2_shuttle}
\end{figure}

Interestingly, the formed NH$_2$CH species is the less stable isomer of methanimine (NH=CH$_2$), a molecule found in different interstellar environments \citep{godfrey1973, Dickens1997}. 
Thus, we studied the isomerization reaction of NH$_2$CH $\rightarrow$ NH=CH$_2$ to check if this transformation is feasible in the ISM. 
We run an NEB calculation (only on Pos1) by considering a water-assisted proton transfer (results shown in Fig. \ref{fig:nh2_shuttle}). 
As it was found for the case of formaldehyde (see above), this process presents an energy barrier, in this case of 14 kJ mol$^{-1}$ considering ZPE corrections. 
This energy barrier is insurmountable at cryogenic interstellar temperatures. 
However, as mentioned above, tunnelling effects can make the reaction kinetically feasible. 
To check this point, we also computed here the crossover temperature, resulting to be 198 K. Thus, this route cannot be excluded, while a more rigorous kinetic treatment including nuclear quantum effects is mandatory to have a definitive answer to this point.

\begin{table*}[ht]
\centering
    \caption{Summary of the reactivity investigated for the $^3$C-OH$_2$ carbon reactive centre with the closed-shell species NH$_3$, CO, CO$_2$ and H$_2$, the atoms H, N and O, and the molecular radicals of OH, CH$_3$ and NH$_2$. 
    The subsequent reactions of the resulting products with H atoms (to be the case) are also shown. 
    The super indices indicate the electronic spin state of the species.}
\label{tab:allreactions}
\begin{tabular}{@{}llll@{}}
\toprule
Reactive species & Reaction                                  & $\Delta$H$_{rx}$(0) & $\Delta$H$^\ddag$(0)       \\ \midrule
NH$_3$              & No reaction                               & -       & -            \\ \midrule
CO               & No reaction                               & -       & -            \\ \midrule
CO$_2$              & No reaction                               & -       & -            \\ \midrule
H$_2$               & H$_2$ \ch{+} $^3$C-OH$_2$   $\rightarrow$ $^3$CH$_2$ \ch{+} H$_2$O                & -187.1  & 70.3         \\ \midrule
H                & H($^2$S) \ch{+} $^3$C-OH$_2$   $\rightarrow$ $^2$HC-OH$_2$                & -358.6  & Barrierless  \\
                 & $^2$HC-OH$_2$ $\rightarrow$ $^2$CH   \ch{+} H$_2$O                     & +149.5  & -            \\
Pos1                 & H($^2$S) \ch{+} $^2$HC-OH$_2$   $\rightarrow$ CH$_3$-OH         & -614.3  & Barrierless  \\
Pos2                 & H($^2$S) \ch{+} $^2$HC-OH$_2$   $\rightarrow$ CH$_2$-OH$_2$ $\xrightarrow{\text{NEB}}$ CH$_3$-OH  & -601.7  & Barrierless* \\
Pos3                &  H($^2$S) \ch{+} $^2$HC-OH$_2$   $\rightarrow$ CH$_2$-OH$_2$ $\xrightarrow{\text{NEB}}$ CH$_3$-OH  &   -638.8 & Barrierless* \\
                 & H($^2$S) \ch{+} $^2$HC-OH$_2$   $\rightarrow$ $^3$CH$_2$ \ch{+} H$_2$O            & -275.8  & Barrierless  \\ \midrule
N                & N($^4$S) \ch{+} $^3$C-OH$_2$   $\rightarrow$ $^2$CN \ch{+} H$_2$O              & -683.9  & Barrierless  \\ \midrule
O                & O($^3$P) \ch{+} $^3$C-OH$_2$   $\rightarrow$ CO \ch{+} H$_2$O               & -958.0  & Barrierless  \\ \midrule
OH               & $^2$OH \ch{+} $^3$C-OH$_2$ $\rightarrow$ $^2$HO-C-OH$_2$                 & -386.2  & Barrierless  \\
                 & H($^2$S) \ch{+} $^2$HO-C-OH$_2$   $\rightarrow$ HO-CH$_2$-OH          & -589.2  & Barrierless  \\
                 & H($^2$S) \ch{+} $^2$HO-C-OH$_2$   $\rightarrow$ HCOH \ch{+} H$_2$O         & -388.8  & Barrierless  \\
                 & HCOH $\rightarrow$ H$_2$CO                               & -170.1  & 7.6          \\ \midrule
CH$_3$              & $^2$CH$_3$ \ch{+} $^3$C-OH$_2$ $\rightarrow$ $^2$CH$_3$-C-OH$_2$                & -328.5  & Barrierless  \\
                 & H($^2$S) \ch{+} $^2$CH$_3$-C-OH$_2$   $\rightarrow$ CH$_3$CH$_2$OH           & -618.6  & Barrierless  \\ \midrule
NH$_2$              & $^2$NH$_2$   \ch{+} $^3$C-OH$_2$ $\rightarrow$ $^2$C-NH$_2$ \ch{+} H$_2$O            & -473.0  & Barrierless  \\
                 & H($^2$S) \ch{+} $^2$C-NH$_2$   $\rightarrow$ NH$_2$-CH                 & -346.0  & Barrierless  \\
                 & $^2$NH$_2$ \ch{+} $^3$C-OH$_2$ $\rightarrow$ $^2$NH$_2$-C-OH$_2$                & -475.5  & Barrierless  \\
                 & H($^2$S) \ch{+} $^2$NH$_2$-C-OH$_2$   $\rightarrow$ NH$_2$-CH \ch{+} H$_2$O       & -270.1  & Barrierless  \\
                 & NH$_2$-CH $\rightarrow$ NH=CH$_2$                           & -99.0   & 14.0         \\ \bottomrule
\end{tabular}
\\
\small{*It presents a potential energy barrier $\leq$ 4 kJ mol$^{-1}$, but when including ZPE-corrections the barrier becomes submerged with respect to the reactants and, accordingly, it is barrierless.}
\end{table*}

\section{Discussion and astrophysical implications} \label{sec:discussion}

\subsection{Carbon reactivity and reaction products}
According to the above-exposed results, the C ($^3$P) atom presents an extraordinary reactivity towards any H$_2$O molecule belonging to interstellar ice mantles, forming the $^3$C-OH$_2$ through the simple coupling of the C atom with the water molecule. 
This very first result supposes an important astrophysical implication: C atoms landing on water-dominated ice mantles do not remain as carbon atoms, but they are locked in the form of $^3$C-OH$_2$ making difficult the occurrence of chemical reactions involving bare C atoms on water icy surfaces.
However, this species keeps the triplet electronic spin state of the initial C atom, which is of fundamental importance for its chemical properties and reactivity. 
A summary of the chemical reactions presented in section \ref{sec:results}, alongside their energetic features (i.e., reaction energies and energy barriers if present), is reported in Table \ref{tab:allreactions}.

The $^3$C-OH$_2$ carbon reactive centre does not present any chemical reactivity when in proximity with the closed-shell species of CO, CO$_2$, NH$_3$ and H$_2$, pointing out that it remains as such once formed on the water ice mantles. However, in the presence of open-shell species like atoms and molecular radicals, it exhibits great chemical activity. As one can see in Table \ref{tab:allreactions}, most of the processes involving open-shell reactants are barrierless with very favourable reaction energies and, accordingly, are highly doable in the ISM conditions. 
This renders the $^3$C-OH$_2$ species a potential trigger for the interstellar chemical diversity and complexity.
However, it mostly leads to the formation of simple molecules, such as CO, CN, and CH$_2$, and only a few iCOMs are formed in this way: methanol, methanediol and ethanol. 
Most of the studied reactions follow a chemical pattern towards the final formation of these compounds. That is, the $^3$C-OH$_2$ carbon reactive centre reacts with the corresponding open-shell species followed by an H addition, which results in the final product, usually involving a spontaneous (i.e., barrierless) water-assisted proton-transfer shuttle from the OH$_2$ moiety to the unsaturated atom. For instance, for the formation of CH$_3$CH$_2$OH, $^3$C-OH$_2$ reacts with $^2$CH$_3$ to form $^2$CH$_3$-COH$_2$, and a subsequent H addition to the central C atom activates a spontaneous proton transfer assisted by water from the OH$_2$ group to the central C atom (see Fig. 8). For those cases in which the water-assisted proton transfer is not spontaneous (i.e., presenting a barrier), which concern the formation of H$_2$CO and NH=CH$_2$ (see Table \ref{tab:allreactions}), the processes are expected to be kinetically feasible. Indeed, the barriers are not high (7.6 and 14.0 kJ mol$^{-1}$), and because of the involvement of the light H atom and the very low temperatures at which these reactions occur, tunnelling effects should dominate the kinetics (as indicated by the crossover temperatures), thereby enabling the occurrence of these reactions. 

It is worth mentioning, however, that all these results hold when the reactive species are in proximity. That is, for the barrierless processes, we observe spontaneous reactivity (e.g., direct chemical bond formation) when the two partners are separated by ca. 2.5 {\AA} and the system is allowed to relax. If we consider that they take place adopting a Langmuir-Hinshelwood mechanism, the reactions are thus limited by the diffusivity of the reactants. This should not be a problem for the atomic species (H, O and N) since they can diffuse on water ice surfaces at very cold temperatures \citep[e.g.[]{Kuwahata2015, shimonishi_adsorption_2018, pezzella_o2_2019,Senevirathne2017-Hdiffusion}. 
However, this is not the case for the CH$_3$, NH$_2$ and OH radical species since, according to their binding energies (ranges of 9-13, 24-37 and 13-44 kJ mol$^{-1}$, respectively \citep{ferrero_binding_2020}, they remain firmly attached to the surface with no chance to diffuse on the water ice surface at 10 K. Thus, the only way for these latter reactions to occur is that the carbon reactive centre forms (i.e., the C atom lands on the ice surface) in the surroundings of these radicals that can be present as icy species (previously formed by photolysis of relatively abundant icy components) or, alternatively, that the radicals form in the vicinity of an already formed $^3$C-OH$_2$ species.

In relation to this aspect of diffusion, it is also worth mentioning that the formation of $^3$C-OH$_2$ implies the chemisorption of C atoms on water ice \citep{shimonishi_adsorption_2018, molpeceres_carbon_2021,Lamberts2022_methane}. 
This goes against the usual vision that C, as an atomic species, is physisorbed and, as such, presents some diffusivity on the surfaces. 
Our results indicate that this is not the case, because C is locked as $^3$C-OH$_2$, which does not show mobility on the water ice surfaces, and should be considered in the future in the astrochemical models.

\subsection{Comparison with previous studies}
Formation of the $^3$C-OH$_2$ species was also found in other works \citep{shimonishi_adsorption_2018, molpeceres_carbon_2021}. 
A comparison with the work of \cite{molpeceres_carbon_2021} is needed to have a clear picture of the possible processes that can happen after the carbon chemisorption on water ice mantles. In their work, a large set of C + H$_2$O(ice) reactions were studied because they sampled more initial positions. 
In the 71\% of the studied cases, they found, like in our work, the direct formation of the $^3$C-OH$_2$ species, while in the remaining cases, they observed either the formation of the $^3$[COH$^-$/H$_3$O$^+$] ion pair (19\%) or the direct formation of $^3$HCOH (10\%), phenomena that are not observed in our (more limited) cases. Subsequently, they studied how the $^3$C-OH$_2$ and the ion pair species can rearrange in $^3$HCOH. They found out that both processes are doable via a water-assisted proton transfer, the former presenting energy barriers that depend on the binding site, that is, very low or even barrierless in shallowed C binding sites, or barriers of 9.7--11.5 kJ mol$^{-1}$ in deeper C binding sites). The ion pair rearrangement was found to have a submerged barrier (e.g barrierless) when including ZPE corrections. From the $^3$HCOH species, they showed that formaldehyde (H$_2$CO) can be formed in a barrierless way but passing through an ISC, involving a change in the spin state (from a triplet electronic state to the singlet one). 
Nevertheless, it is reasonable to think that the $^3$C-OH$_2$, a relatively stable compound, can also undergo the reaction with the open-shell species in its proximity, forming the variety of compounds shown in this work, can be regarded as actual competitive processes to formaldehyde formation due to the restructuring of the $^3$C-OH$_2$ carbon reactive centre. The possible outcomes will highly depend on the binding sites of the C atom and the surrounding structure of the ice. 

\subsection{Proton transfer process}
An important aspect that allows the formation of the final products is the occurrence of water-assisted proton transfers. As highlighted in different works dealing with these processes \citep{molpeceres_carbon_2021, rimola_can_2018, perrero_non-energetic_2022}, this mechanism is efficient only if water molecules surrounding the carbon atom have a favourable orientation for the proton transfer, that is, the proton to be transferred and the water molecules have to be connected by hydrogen bonds (H-bonds) that permit the proton shuttle by linking the initial position to the final one. In our reaction, this proper orientation is facilitated by the negative net charge of the C atom in the $^2$H$_2$C-OH$_2$ species (as provided by the Löwdin atomic charges, see above), which converts the carbon atom as a good H-bond acceptor, this way forming part of the H-bond connectivity and receiving the shuttled proton. 
It is worth mentioning that water-assisted mechanisms have been studied in other interstellar reactions, in which in some cases water molecules infer a strong catalytic effect (by reducing the energy barriers or even rendering them as spontaneous, like in this work \citep{molpeceres_carbon_2021, rimola_can_2018, perrero_non-energetic_2022, rimola2010deep}), while in others the transfer exhibits high potential energy barriers \citep{enrique-romero_reactivity_2019, enrique-romero_revisiting_2020}. The occurrence of one event or the other depends on the chemical nature of the H atom, that is, whether it has a proton-like or a radical-like character. In the former, the water-assisted mechanism is efficient, the transfer presenting low energy barriers, while in the latter, the process presents high energy barriers. In water-assisted processes involving proton-like atoms, the exchanged H atoms have positive net charges (and hence they are H$^+$ (proton)-like atoms) and accordingly, the transfer occurs through a set of breaking/making chemical bonds with more electronegative atoms (in our reactions, through the O atoms of the water molecules and the C atom with a negative net charge), in which transient H$_3$O$^{+}$ species are formed along the proton shuttle. These H$_3$O$^+$ are stabilized by the interaction with the surrounding water molecules, which makes the water-assisted proton transfer energetically favourable. In contrast, in water-assisted processes involving H radical-like atoms, during the H shuttle, the radical H$_3$O$^{\cdot}$ forms as a transient species, which is not stable (it tends to give H$_2$O + H$^{\cdot}$, see e.g. \citep{rimola_interaction_2021}) and hence that, for these cases, the water-assisted mechanism presents high energy barriers.

\subsection{Dependence on the substrate local structure}
Another message that emerges from this work is the relevance of the local structure of the ice in the surroundings where the $^3$C-OH$_2$ forms. In this work, three different positions have been considered (Pos1, Pos2 and Pos3) and according to our results, Pos3 presents different environmental effects with respect to Pos1 and Pos2. Obviously, robust statistics of the possible outcomes of the reactions cannot be reached with the three studied cases only. However, the molecular pictures provided here indicate that the icy water interactions with $^3$C-OH$_2$ are essential in modulating its chemical properties and reactivity. That is, in Pos3, the C atom of $^3$C-OH$_2$ is engaged in three hydrogen bonds, whereas in Pos1 and Pos2 in only two (see Fig. \ref{fig:c_ads} for comparison). Accordingly, $^3$C-OH$_2$ in Pos3 $^3$C-OH$_2$ presents an enhanced stabilization with respect to those in Pos1 and Pos2. Hence, it was found that the addition of the NH$_2$ radical on the $^3$C-OH$_2$ in Pos1 and Pos2 results in the cleavage of the C-O bond (releasing the original H$_2$O molecule), while in Pos3 the $^3$C-OH$_2$ species remains stable. Therefore, the structural environment around the carbon reactive centre is an important aspect to take into consideration to understand its reactivity, since different routes can dominate according to the boundary conditions.

\subsection{Methane formation}
Finally, this work also gives an atomistic interpretation of experiments performed in laboratories relative to the reactivity of atomic C in the presence of water ice surfaces. As mentioned in the Introduction, \citet{QASIM_CH4} found that methane (CH$_4$) forms when C- and H-atoms, and H$_2$O molecules were co-deposited at cryogenic temperatures. Although a priori our results go against these findings (because C on H$_2$O ice is locked in the form of $^3$C-OH$_2$), this is not the case, and we provide the following mechanistic proposal towards CH$_4$ formation that reconciles theory and experiment. According to our results, a first H addition to $^3$C-OH$_2$ gives $^2$HC-OH$_2$, and a second H addition forms CH$_2$-OH$_2$, which converts into CH$_3$OH through a spontaneous water-assisted proton transfer. This is true if the second H addition is performed in a total singlet electronic state, namely, the $^2$HC-OH$_2$ and the H ($^2$S) present opposite electronic spins. However, if the reaction occurs in a total triplet electronic state (i.e., with the two partners presenting unpaired electrons having the same spin), the reaction gives $^3$CH$_2$-OH$_2$, which transforms spontaneously into $^3$CH$_2$ + H$_2$O, that is, forming methylene ($^3$CH$_2$), and releasing the original water. This process has been found to be barrierless and has a reaction energy of -245 kJ mol$^{-1}$ (see Table \ref{tab:allreactions}) and, thus, it is feasible and competing with the singlet reaction. Accordingly, in the experiments, it is likely that atomic C condensates with H$_2$O, forming the $^3$C-OH$_2$ carbon reactive centre, the hydrogenation of which gives rise first $^3$CH$_2$ and finally CH$_4$. 
Unfortunately, experiments did not search for methanol so that we cannot say if they found it or not.

Our results are also in agreement with the work of \citet{Lamberts2022_methane} on the reactivity of H$_2$ with C- and CH$_n$ species to form CH$_4$. We have computed the reaction of H$_2$ + $^3$C-OH$_2$ $\rightarrow$ $^3$CH$_2$ + H$_2$O on our cluster model and it presents an energy barrier of 70 kJ mol$^{-1}$. This value is very close to the value found in \citet{Lamberts2022_methane} for the reaction in the presence of a (H$_2$O)$_3$ minimal cluster model. These energy barrier values are extremely high to occur in the ISM, even if tunneling could dominate. Thus, this path to form CH$_4$ can be excluded as concluded in \citet{Lamberts2022_methane}. A likely possibility already pointed out in \citet{Lamberts2022_methane} work, is that the $^2$CH-OH$_2$ species is first formed, and reaction with H$_2$ gives rise to CH$_3$ and H$_2$O, the former species giving rise to CH$_4$ upon hydrogenation. 

\subsection{Astrophysical implications}

Since the pioneering work by \cite{Garrod2006-iCOMmodel}, the astrochemical models have privileged the scheme of radical-radical combination and/or hydrogenation of molecules on the grain-surfaces for the synthesis of iCOMs.
The former reaction scheme postulates that radicals diffuse over the grain-surfaces upon the increase of the dust grain temperature caused by the birth of a protostar.
However, this scheme has been challenged by the detection of iCOMs in protostellar shocks \citep[e.g.][]{Arce2008, Codella2009-B1ch3cn, Lefloch2017, Codella2017-B1formamide, simone_seeds_2020} and cold objects \citep[e.g.][]{Bacmann2012, Cernicharo2012, Vastel2014-PSCiCOM, Scibelli2020-PSC}, where the dust temperature has never increased and, consequently, radicals could not migrate.
Additionally, QM calculations show that the radical-radical combination does not necessarily lead to the iCOM formation \citep{enrique-romero_theoretical_2021, enrique-romero_quantum_2022}.

Alternative possibilities have been proposed, such as non-diffusive processes occurring on the grain-surfaces \citep{Jin2020-iCOMnondiff, Garrod2022-iCOMnondiff} and the formation of iCOMs in the gas-phase \citep{vasyunin_reactive_2013, Balucani2015-PSCiCOM, Skouteris2017-formamide, Skouteris2018-ethanoltree}.

Another possibility is the one studied in this work, namely that gaseous species landing instantaneously react with the frozen molecules of the grain mantles.
Given their abundances, water and CO are the two major possible ice-molecules.
Previous works have shown that this is a viable path to synthesise formamide and ethanol on H$_2$O-rich ice \citep{rimola_can_2018, perrero_non-energetic_2022} and acetaldehyde and ethanol on CO-rich ice \citep{Ferrero2023-C+CO}.
In this work, we have examined the possible formation of iCOMs from gaseous C atoms landing on H$_2$O-rich ices, which gives rise to the formation of the very reactive, and non-diffusive, species $^3$C-OH$_2$ \citep[see also][]{shimonishi_adsorption_2018, molpeceres_carbon_2021}.
In turn, this species could react with landing or nearby atoms and molecules.
Here we considered the most abundant of them and found that some reactions lead to small species while only methanol, methanediol and ethanol can be formed in this way.

Therefore, another process beyond CO hydrogenation can produce methanol on the grain-surfaces, started by C atoms landing on H$_2$O-rich ice. 
This is a potentially important contribution to the methanol formation in regions where C and not CO is abundant, notably the Photo-Dissociation Regions (PDRs) and in the early stages of the chemical evolution of molecular clouds from diffuse atomic ones.
Gaseous methanol has indeed been observed in these regions \citep[see, e.g., the discussion in][]{Bouvier2020-methanolPDR}.
We have to emphasise that, assuming that the process here found to form frozen methanol is the dominant one in CO-poor/C-rich regions, some non-thermal mechanism must be at work to inject it into the gas-phase. In this sense, three mechanisms are evoked in the literature: photo-desorption, reactive desorption and cosmic-ray induced desorption.
However, all of them seem to have drawbacks.
UV-induced photo-desorption does break the methanol in small pieces \citep{Bertin2016-UVphotodes}, reactive desorption does not seem to be efficient \citep{pantaleone_chemical_2020}, as well as cosmic-ray induced desorption \citep{Wakelam2021-CRdes}.

Methanediol is not detected yet in the ISM \citep{mcguire2021}, perhaps because the frequencies of its rotational transitions are not available in the two astronomical databases CDMS \citep[\url{https://cdms.astro.uni-koeln.de/cgi-bin/cdmssearch}:][]{Endres2016-CDMS} and JPL \citep[\url{https://spec.jpl.nasa.gov/ftp/pub/catalog/catform.html}:][]{Pickett1998-JPL}.
Probably, its detection (even with JWST) on the solid phase is unlikely because of the a priori low abundance and also because the frequencies are likely in a heavily crowded region of the spectrum.
It will be worth searching this molecule on the gas-phase: if a non-thermal mechanism injects methanol, likely also some methanediol will be gaseous.

Finally, our work shows that ethanol can be synthesised on the H$_2$O-rich ice grain-surfaces in CO-poor/C-rich regions, i.e. in PDRs and in the early stages of the chemical evolution of molecular clouds from diffuse atomic ones.
\cite{perrero_non-energetic_2022} have already shown that ethanol can be formed on the grain-surfaces in H$_2$O-rich ices by the reaction of CCH with an icy water molecule, followed by hydrogenation.
In addition, \cite{enrique-romero_quantum_2022} has shown that the coupling of CH$_3$ and CH$_2$OH on the grain-surfaces may also lead to ethanol.
Thus, there are at least three different grain-surface paths that lead to the formation of ethanol. 
In the astrochemical gas-phase reaction networks, ethanol can be formed either by the electron recombination of protonated ethanol or the reaction of acetone with H$_3^+$ (\url{http://udfa.ajmarkwick.net/}).
Considering the gaseous abundances of C, CCH, protonated ethanol and acetone and the likely frozen abundances of CH$_3$ and CH$_2$OH, the most efficient way to form ethanol is likely on the grain surfaces via the reactions involving C and CCH, whose abundances can reach a maximum of about $10^{-4}$ and $10^{-8}$ (with respect to H-nuclei), respectively.
Moreover, it is possible that the path involving carbon atoms, studied here, is the dominant way to form ethanol.
Remarkably, ethanol is the only iCOMs in addition to methanol which may have been detected in the solid form \citep{Yang2022-JWST, McClure2023}.
Also remarkably, ethanol is supposed to be the mother of some iconic iCOMs, glycolaldehyde and acetaldehyde \citep{Skouteris2018-ethanoltree, Vazart2020-acetaldehyde}, once injected into the gas-phase.

\section{Conclusions} \label{sec:conclusions}

In this work, the reactivity of atomic C on water ice has 
been studied by means of quantum chemical computations. 
It is shown that the C atom in its ground state ($^3$P) is reactive upon water ice adsorption forming the $^3$C-OH$_2$ species, which remains anchored to the water ice. 
Then, possible subsequent reactions with closed-shell species (i.e., CO, CO$_2$, NH$_3$ and H$_2$), atoms (i.e., H($^1$S), N($^4$S) and O($^3$P)) and radicals (i.e., OH, NH$_2$ and CH$_3$) have been considered. 

It has been found that $^3$C-OH$_2$ does not present reactivity with the closed-shell species, but it is highly reactive with the atoms and molecular radicals.
Some reactions form small molecules, such as CN and CO.
Others lead to the formation of three iCOMs spontaneously, in a barrierless process: methanol (CH$_3$OH), methanediol (HOCH$_2$OH), and ethanol (CH$_3$CH$_2$OH). 
In addition, the formation of H$_2$CO and NH=CH$_2$ exhibit energy barriers not surmountable at cryogenic temperatures, but they might be formed via tunneling, as indicated by the respective calculated crossover temperatures. 
In view of the low energy requirements of these chemical reactions, the processes are limited by diffusion, which for the molecular radicals is an important issue due to their low diffusivity. Accordingly, the reactions are feasible if the two partners are in proximity in cold environments. 

In most of the formations of these species, water ice actively participates in the reaction thanks to water-assisted proton transfers. 
They happen in a barrierless way at low ISM temperatures but only work if the icy water components present a suitable hydrogen bond connection that allows the proton exchange. 

The work also shows the role of icy grains as concentrators of C in an activated form, since the $^3$C-OH$_2$ species is highly reactive with open-shell species. To the best of our knowledge, this feature is not particularly developed in astrochemical models as atomic C is usually considered to be physisorbed on ice surfaces. This also affects the diffusion properties of C, since it is completely hindered, at variance to what would occur if it were physisorbed.

From the astrochemical context, according to our results, it seems that CH$_4$ cannot form by multiple hydrogenations of atomic C on water ice mantles because it is locked in the form of $^3$C-OH$_2$. 
However, the great reactivity of this species allows a likely final formation of CH$_4$ (as observed in the experiments of \citep{QASIM_CH4}. 
This is achieved, once the $^3$C-OH$_2$ species is formed, as its subsequent hydrogenation (i.e., reaction with H ($^2$S)) gives rise to first the $^2$CH-OH$_2$ intermediate and then $^3$CH$_2$-OH$_2$ (if the reaction takes place in a triplet total electronic spin state), which decomposes spontaneously into $^3$CH$_2$ + H$_2$O. 
Two H additions on the generated $^3$CH$_2$ result in the formation of CH$_4$. 
Our results also agree with the work of \citet{Lamberts2022_methane}, which rules out the reactivity of molecular hydrogen (H$_2$) with C as a reactive channel towards CH$_4$ formation. 
In our case, the reaction of H$_2$ with the $^3$C-OH$_2$ to give $^3$CH$_2$ + H$_2$O presents a huge energy barrier insurmountable at the ISM conditions. 

Finally, this work is particularly relevant for the formation of methanol and ethanol in CO-poor/C-rich regions, such as PDRs or in the early chemical evolution of molecular clouds.
When taking into account the different proposed routes to form ethanol and the abundances of the reactants, we showed that the one involving atomic carbon and water-ice molecules, as studied here, could be the dominant one.


\section{acknowledgments}
This project has received funding from the Marie Sklodowska-Curie for the project ``Astro-Chemical Origins” (ACO), grant agreement No 811312, and within the European Union’s Horizon 2020 research and innovation program from the European Research Council (ERC) for the projects ``The Dawn of Organic Chemistry” (DOC), grant agreement No 741002, and ``Quantum Chemistry on Interstellar Grains” (QUANTUMGRAIN), grant agreement No 865657. MICIN (project
PID2021-126427NB-I00) is also acknowledged.

\bibliography{bibliography.bib}

\begin{thebibliography}{}
\expandafter\ifx\csname natexlab\endcsname\relax\def\natexlab#1{#1}\fi
\providecommand{\url}[1]{\href{#1}{#1}}
\providecommand{\dodoi}[1]{doi:~\href{http://doi.org/#1}{\nolinkurl{#1}}}
\providecommand{\doeprint}[1]{\href{http://ascl.net/#1}{\nolinkurl{http://ascl.net/#1}}}
\providecommand{\doarXiv}[1]{\href{https://arxiv.org/abs/#1}{\nolinkurl{https://arxiv.org/abs/#1}}}

\bibitem[{{Allen} \& {Robinson}(1975)}]{Allen1975-surfacechemistry}
{Allen}, M., \& {Robinson}, G.~W. 1975, \apj, 195, 81, \dodoi{10.1086/153306}

\bibitem[{{Andersson} {et~al.}(2011){Andersson}, {Goumans}, \&
  {Arnaldsson}}]{andersson_tun}
{Andersson}, S., {Goumans}, T.~P.~M., \& {Arnaldsson}, A. 2011, Chemical
  Physics Letters, 513, 31, \dodoi{10.1016/j.cplett.2011.07.073}

\bibitem[{{Arce} {et~al.}(2008){Arce}, {Santiago-Garc{\'\i}a}, {J{\o}rgensen},
  {Tafalla}, \& {Bachiller}}]{Arce2008}
{Arce}, H.~G., {Santiago-Garc{\'\i}a}, J., {J{\o}rgensen}, J.~K., {Tafalla},
  M., \& {Bachiller}, R. 2008, \apjl, 681, L21, \dodoi{10.1086/590110}

\bibitem[{{Bacmann} {et~al.}(2012){Bacmann}, {Taquet}, {Faure}, {Kahane}, \&
  {Ceccarelli}}]{Bacmann2012}
{Bacmann}, A., {Taquet}, V., {Faure}, A., {Kahane}, C., \& {Ceccarelli}, C.
  2012, \aap, 541, L12, \dodoi{10.1051/0004-6361/201219207}

\bibitem[{{Balucani} {et~al.}(2015){Balucani}, {Ceccarelli}, \&
  {Taquet}}]{Balucani2015-PSCiCOM}
{Balucani}, N., {Ceccarelli}, C., \& {Taquet}, V. 2015, \mnras, 449, L16,
  \dodoi{10.1093/mnrasl/slv009}

\bibitem[{{Bernstein} {et~al.}(1995){Bernstein}, {Sandford}, {Allamandola},
  {Chang}, \& {Scharberg}}]{Bernstein1995ApJ...454..327B}
{Bernstein}, M.~P., {Sandford}, S.~A., {Allamandola}, L.~J., {Chang}, S., \&
  {Scharberg}, M.~A. 1995, \apj, 454, 327, \dodoi{10.1086/176485}

\bibitem[{{Bernstein} {et~al.}(1999){Bernstein}, {Sandford}, {Allamandola},
  {Gillette}, {Clemett}, \& {Zare}}]{Bernstein1999Sci...283.1135B}
{Bernstein}, M.~P., {Sandford}, S.~A., {Allamandola}, L.~J., {et~al.} 1999,
  Science, 283, 1135, \dodoi{10.1126/science.283.5405.1135}

\bibitem[{{Bertin} {et~al.}(2016){Bertin}, {Romanzin}, {Doronin}, {Philippe},
  {Jeseck}, {Ligterink}, {Linnartz}, {Michaut}, \&
  {Fillion}}]{Bertin2016-UVphotodes}
{Bertin}, M., {Romanzin}, C., {Doronin}, M., {et~al.} 2016, \apjl, 817, L12,
  \dodoi{10.3847/2041-8205/817/2/L12}

\bibitem[{{Blake} {et~al.}(1987){Blake}, {Sutton}, {Masson}, \&
  {Phillips}}]{Blake1987-orion}
{Blake}, G.~A., {Sutton}, E.~C., {Masson}, C.~R., \& {Phillips}, T.~G. 1987,
  \apj, 315, 621, \dodoi{10.1086/165165}

\bibitem[{Boogert {et~al.}(2015)Boogert, Gerakines, \&
  Whittet}]{boogert_observations_2015}
Boogert, A.~A., Gerakines, P.~A., \& Whittet, D.~C. 2015, \araa, 53, 541,
  \dodoi{10.1146/annurev-astro-082214-122348}

\bibitem[{{Bouvier} {et~al.}(2020){Bouvier}, {L{\'o}pez-Sepulcre},
  {Ceccarelli}, {Kahane}, {Imai}, {Sakai}, {Yamamoto}, \&
  {Dagdigian}}]{Bouvier2020-methanolPDR}
{Bouvier}, M., {L{\'o}pez-Sepulcre}, A., {Ceccarelli}, C., {et~al.} 2020, \aap,
  636, A19, \dodoi{10.1051/0004-6361/201937164}

\bibitem[{{Cazaux} {et~al.}(2003){Cazaux}, {Tielens}, {Ceccarelli}, {Castets},
  {Wakelam}, {Caux}, {Parise}, \& {Teyssier}}]{Cazaux2003}
{Cazaux}, S., {Tielens}, A.~G.~G.~M., {Ceccarelli}, C., {et~al.} 2003, \apjl,
  593, L51, \dodoi{10.1086/378038}

\bibitem[{{Ceccarelli} {et~al.}(2000){Ceccarelli}, {Loinard}, {Castets},
  {Faure}, \& {Lefloch}}]{Ceccarelli2000-glycine}
{Ceccarelli}, C., {Loinard}, L., {Castets}, A., {Faure}, A., \& {Lefloch}, B.
  2000, \aap, 362, 1122

\bibitem[{Ceccarelli {et~al.}(2017)Ceccarelli, Caselli, Fontani, Neri,
  López-Sepulcre, Codella, Feng, Jiménez-Serra, Lefloch, Pineda, Vastel,
  Alves, Bachiller, Balucani, Bianchi, Bizzocchi, Bottinelli, Caux,
  Chacón-Tanarro, Choudhury, Coutens, Dulieu, Favre, Hily-Blant, Holdship,
  Kahane, Al-Edhari, Laas, Ospina, Oya, Podio, Pon, Punanova, Quenard, Rimola,
  Sakai, Sims, Spezzano, Taquet, Testi, Theulé, Ugliengo, Vasyunin, Viti,
  Wiesenfeld, \& Yamamoto}]{ceccarelli_seeds_2017}
Ceccarelli, C., Caselli, P., Fontani, F., {et~al.} 2017, ApJ, 850, 176,
  \dodoi{10.3847/1538-4357/aa961d}

\bibitem[{{Ceccarelli} {et~al.}(2022){Ceccarelli}, {Codella}, {Balucani},
  {Bockel{\'e}e-Morvan}, {Herbst}, {Vastel}, {Caselli}, {Favre}, {Lefloch}, \&
  {{\"O}berg}}]{Ceccarelli2022-PP7}
{Ceccarelli}, C., {Codella}, C., {Balucani}, N., {et~al.} 2022,
  arXiv:2206.13270, \dodoi{10.48550/arXiv.2206.13270}

\bibitem[{{Cernicharo} {et~al.}(2012){Cernicharo}, {Marcelino}, {Roueff},
  {Gerin}, {Jim{\'e}nez-Escobar}, \& {Mu{\~n}oz Caro}}]{Cernicharo2012}
{Cernicharo}, J., {Marcelino}, N., {Roueff}, E., {et~al.} 2012, \apjl, 759,
  L43, \dodoi{10.1088/2041-8205/759/2/L43}

\bibitem[{{Codella} {et~al.}(2009){Codella}, {Benedettini}, {Beltr{\'a}n},
  {Gueth}, {Viti}, {Bachiller}, {Tafalla}, {Cabrit}, {Fuente}, \&
  {Lefloch}}]{Codella2009-B1ch3cn}
{Codella}, C., {Benedettini}, M., {Beltr{\'a}n}, M.~T., {et~al.} 2009, \aap,
  507, L25, \dodoi{10.1051/0004-6361/200913340}

\bibitem[{{Codella} {et~al.}(2017){Codella}, {Ceccarelli}, {Caselli},
  {Balucani}, {Barone}, {Fontani}, {Lefloch}, {Podio}, {Viti}, {Feng},
  {Bachiller}, {Bianchi}, {Dulieu}, {Jim{\'e}nez-Serra}, {Holdship}, {Neri},
  {Pineda}, {Pon}, {Sims}, {Spezzano}, {Vasyunin}, {Alves}, {Bizzocchi},
  {Bottinelli}, {Caux}, {Chac{\'o}n-Tanarro}, {Choudhury}, {Coutens}, {Favre},
  {Hily-Blant}, {Kahane}, {Jaber Al-Edhari}, {Laas}, {L{\'o}pez-Sepulcre},
  {Ospina}, {Oya}, {Punanova}, {Puzzarini}, {Quenard}, {Rimola}, {Sakai},
  {Skouteris}, {Taquet}, {Testi}, {Theul{\'e}}, {Ugliengo}, {Vastel}, {Vazart},
  {Wiesenfeld}, \& {Yamamoto}}]{Codella2017-B1formamide}
{Codella}, C., {Ceccarelli}, C., {Caselli}, P., {et~al.} 2017, \aap, 605, L3,
  \dodoi{10.1051/0004-6361/201731249}

\bibitem[{Cuppen {et~al.}(2017)Cuppen, Walsh, Lamberts, Semenov, Garrod,
  Penteado, \& Ioppolo}]{cuppen_grain_2017}
Cuppen, H.~M., Walsh, C., Lamberts, T., {et~al.} 2017, Space Sci Rev, 212, 1,
  \dodoi{10.1007/s11214-016-0319-3}

\bibitem[{De~Simone {et~al.}(2020)De~Simone, Codella, Ceccarelli,
  López-Sepulcre, Witzel, Neri, Balucani, Caselli, Favre, Fontani, Lefloch,
  Ospina-Zamudio, Pineda, \& Taquet}]{simone_seeds_2020}
De~Simone, M., Codella, C., Ceccarelli, C., {et~al.} 2020, A\&A, 640, A75,
  \dodoi{10.1051/0004-6361/201937004}

\bibitem[{{Dickens} {et~al.}(1997){Dickens}, {Irvine}, {DeVries}, \&
  {Ohishi}}]{Dickens1997}
{Dickens}, J.~E., {Irvine}, W.~M., {DeVries}, C.~H., \& {Ohishi}, M. 1997,
  \apj, 479, 307, \dodoi{10.1086/303884}

\bibitem[{Drechsel-Grau \& Marx(2014)}]{drechsel-grau_quantum_2014}
Drechsel-Grau, C., \& Marx, D. 2014, Phys. Rev. Lett., 112, 148302,
  \dodoi{10.1103/PhysRevLett.112.148302}

\bibitem[{{Duley} {et~al.}(1978){Duley}, {Millar}, \& {Williams}}]{duley1978}
{Duley}, W.~W., {Millar}, T.~J., \& {Williams}, D.~A. 1978, \mnras, 185, 915,
  \dodoi{10.1093/mnras/185.4.915}

\bibitem[{{Endres} {et~al.}(2016){Endres}, {Schlemmer}, {Schilke}, {Stutzki},
  \& {M{\"u}ller}}]{Endres2016-CDMS}
{Endres}, C.~P., {Schlemmer}, S., {Schilke}, P., {Stutzki}, J., \&
  {M{\"u}ller}, H. S.~P. 2016, J. Mol. Spectrosc., 327, 95,
  \dodoi{10.1016/j.jms.2016.03.005}

\bibitem[{Enrique-Romero {et~al.}(2021)Enrique-Romero, Ceccarelli, Rimola,
  Skouteris, Balucani, \& Ugliengo}]{enrique-romero_theoretical_2021}
Enrique-Romero, J., Ceccarelli, C., Rimola, A., {et~al.} 2021, A\&A, 655, A9,
  \dodoi{10.1051/0004-6361/202141531}

\bibitem[{Enrique-Romero {et~al.}(2019)Enrique-Romero, Rimola, Ceccarelli,
  Ugliengo, Balucani, \& Skouteris}]{enrique-romero_reactivity_2019}
Enrique-Romero, J., Rimola, A., Ceccarelli, C., {et~al.} 2019, ACS Earth Space
  Chem., 3, 2158, \dodoi{10.1021/acsearthspacechem.9b00156}

\bibitem[{Enrique-Romero {et~al.}(2022)Enrique-Romero, Rimola, Ceccarelli,
  Ugliengo, Balucani, \& Skouteris}]{enrique-romero_quantum_2022}
---. 2022, ApJS, 259, 39, \dodoi{10.3847/1538-4365/ac480e}

\bibitem[{Enrique-Romero {et~al.}(2020)Enrique-Romero, Álvarez Barcia, Kolb,
  Rimola, Ceccarelli, Balucani, Meisner, Ugliengo, Lamberts, \&
  Kästner}]{enrique-romero_revisiting_2020}
Enrique-Romero, J., Álvarez Barcia, S., Kolb, F.~J., {et~al.} 2020, MNRAS,
  493, 2523, \dodoi{10.1093/mnras/staa484}

\bibitem[{{Favre} {et~al.}(2018){Favre}, {Fedele}, {Semenov}, {Parfenov},
  {Codella}, {Ceccarelli}, {Bergin}, {Chapillon}, {Testi}, {Hersant},
  {Lefloch}, {Fontani}, {Blake}, {Cleeves}, {Qi}, {Schwarz}, \&
  {Taquet}}]{Favre2018}
{Favre}, C., {Fedele}, D., {Semenov}, D., {et~al.} 2018, \apjl, 862, L2,
  \dodoi{10.3847/2041-8213/aad046}

\bibitem[{Fedoseev {et~al.}(2022)Fedoseev, Qasim, Chuang, Ioppolo, Lamberts,
  Dishoeck, \& Linnartz}]{fedoseev_hydrogenation_2022}
Fedoseev, G., Qasim, D., Chuang, K.-J., {et~al.} 2022, ApJ, 924, 110,
  \dodoi{10.3847/1538-4357/ac3834}

\bibitem[{Fermann \& Auerbach(2000)}]{fermann2000modeling}
Fermann, J.~T., \& Auerbach, S. 2000, J Chem. Phys., 112, 6787,
  \dodoi{https://doi.org/10.1063/1.481318}

\bibitem[{{Ferrero} {et~al.}(2023{\natexlab{a}}){Ferrero}, {Ceccarelli},
  {Ugliengo}, {Sodupe}, \& {Rimola}}]{Ferrero2023-C+CO}
{Ferrero}, S., {Ceccarelli}, C., {Ugliengo}, P., {Sodupe}, M., \& {Rimola}, A.
  2023{\natexlab{a}}, \apj, 951, 150,
  \dodoi{https://doi.org/10.3847/1538-4357/acd192}

\bibitem[{{Ferrero} {et~al.}(2023{\natexlab{b}}){Ferrero}, {Pantaleone},
  {Ceccarelli}, {Ugliengo}, {Sodupe}, \& {Rimola}}]{FerreroNH3-2023}
{Ferrero}, S., {Pantaleone}, S., {Ceccarelli}, C., {et~al.} 2023{\natexlab{b}},
  \apj, 944, 142, \dodoi{10.3847/1538-4357/acae8e}

\bibitem[{Ferrero {et~al.}(2020)Ferrero, Zamirri, Ceccarelli, Witzel, Rimola,
  \& Ugliengo}]{ferrero_binding_2020}
Ferrero, S., Zamirri, L., Ceccarelli, C., {et~al.} 2020, ApJ, 904, 11,
  \dodoi{10.3847/1538-4357/abb953}

\bibitem[{{Garrod}(2013)}]{Garrod013-glycine}
{Garrod}, R.~T. 2013, \apj, 765, 60, \dodoi{10.1088/0004-637X/765/1/60}

\bibitem[{{Garrod} \& {Herbst}(2006)}]{Garrod2006-iCOMmodel}
{Garrod}, R.~T., \& {Herbst}, E. 2006, \aap, 457, 927,
  \dodoi{10.1051/0004-6361:20065560}

\bibitem[{{Garrod} {et~al.}(2022){Garrod}, {Jin}, {Matis}, {Jones}, {Willis},
  \& {Herbst}}]{Garrod2022-iCOMnondiff}
{Garrod}, R.~T., {Jin}, M., {Matis}, K.~A., {et~al.} 2022, \apjs, 259, 1,
  \dodoi{10.3847/1538-4365/ac3131}

\bibitem[{{Godfrey} {et~al.}(1973){Godfrey}, {Brown}, {Robinson}, \&
  {Sinclair}}]{godfrey1973}
{Godfrey}, P.~D., {Brown}, R.~D., {Robinson}, B.~J., \& {Sinclair}, M.~W. 1973,
  \aplett, 13, 119

\bibitem[{Goerigk {et~al.}(2017)Goerigk, Hansen, Bauer, Ehrlich, Najibi, \&
  Grimme}]{goerigk_look_2017}
Goerigk, L., Hansen, A., Bauer, C., {et~al.} 2017, Phys. Chem. Chem. Phys., 19,
  32184, \dodoi{10.1039/C7CP04913G}

\bibitem[{Hama \& Watanabe(2013)}]{hama_surface_2013}
Hama, T., \& Watanabe, N. 2013, Chem. Rev., 113, 8783,
  \dodoi{10.1021/cr4000978}

\bibitem[{{Henkel} {et~al.}(1987){Henkel}, {Jacq}, {Mauersberger}, {Menten}, \&
  {Steppe}}]{Henkel1987-CH3OHextragal}
{Henkel}, C., {Jacq}, T., {Mauersberger}, R., {Menten}, K.~M., \& {Steppe}, H.
  1987, \aap, 188, L1

\bibitem[{Henning(2010)}]{henning_cosmic_2010}
Henning, T. 2010, \araa, 48, 21, \dodoi{10.1146/annurev-astro-081309-130815}

\bibitem[{Henning \& Krasnokutski(2019)}]{henning_experimental_2019}
Henning, T.~K., \& Krasnokutski, S.~A. 2019, Nat Astron, 3, 568,
  \dodoi{10.1038/s41550-019-0729-8}

\bibitem[{Herbst \& van Dishoeck(2009)}]{herbst_complex_2009}
Herbst, E., \& van Dishoeck, E.~F. 2009, \araa, 47, 427,
  \dodoi{10.1146/annurev-astro-082708-101654}

\bibitem[{Humphrey {et~al.}(1996)Humphrey, Dalke, \&
  Schulten}]{humphrey_vmd_1996}
Humphrey, W., Dalke, A., \& Schulten, K. 1996, J. Mol. Graph., 14, 33,
  \dodoi{10.1016/0263-7855(96)00018-5}

\bibitem[{Hwang {et~al.}(1999)Hwang, Mebel, \& Wang}]{HWANG1999143}
Hwang, D.-Y., Mebel, A.~M., \& Wang, B.-C. 1999, Chem. Phys., 244, 143,
  \dodoi{https://doi.org/10.1016/S0301-0104(99)00156-1}

\bibitem[{{Ilee} {et~al.}(2021){Ilee}, {Walsh}, {Booth}, {Aikawa}, {Andrews},
  {Bae}, {Bergin}, {Bergner}, {Bosman}, {Cataldi}, {Cleeves}, {Czekala},
  {Guzm{\'a}n}, {Huang}, {Law}, {Le Gal}, {Loomis}, {M{\'e}nard}, {Nomura},
  {{\"O}berg}, {Qi}, {Schwarz}, {Teague}, {Tsukagoshi}, {Wilner}, {Yamato}, \&
  {Zhang}}]{Ilee2021-iCOMdisks}
{Ilee}, J.~D., {Walsh}, C., {Booth}, A.~S., {et~al.} 2021, \apjs, 257, 9,
  \dodoi{10.3847/1538-4365/ac1441}

\bibitem[{{Iqbal} \& {Wakelam}(2018)}]{Iqbal2018}
{Iqbal}, W., \& {Wakelam}, V. 2018, \aap, 615, A20,
  \dodoi{10.1051/0004-6361/201732486}

\bibitem[{{Jin} \& {Garrod}(2020)}]{Jin2020-iCOMnondiff}
{Jin}, M., \& {Garrod}, R.~T. 2020, \apjs, 249, 26,
  \dodoi{10.3847/1538-4365/ab9ec8}

\bibitem[{Jones {et~al.}(2013)Jones, Fanciullo, Köhler, Verstraete, Guillet,
  Bocchio, \& Ysard}]{jones_evolution_2013}
Jones, A.~P., Fanciullo, L., Köhler, M., {et~al.} 2013, A\&A, 558, A62,
  \dodoi{10.1051/0004-6361/201321686}

\bibitem[{Jones {et~al.}(2017)Jones, Köhler, Ysard, Bocchio, \&
  Verstraete}]{jones_global_2017}
Jones, A.~P., Köhler, M., Ysard, N., Bocchio, M., \& Verstraete, L. 2017,
  A\&A, 602, A46, \dodoi{10.1051/0004-6361/201630225}

\bibitem[{{J{\o}rgensen} {et~al.}(2020){J{\o}rgensen}, {Belloche}, \&
  {Garrod}}]{Jorgensen2020-ARA&A}
{J{\o}rgensen}, J.~K., {Belloche}, A., \& {Garrod}, R.~T. 2020, \araa, 58, 727,
  \dodoi{10.1146/annurev-astro-032620-021927}

\bibitem[{Krasnokutski {et~al.}(2022)Krasnokutski, Chuang, Jäger, Ueberschaar,
  \& Henning}]{krasnokutski_pathway_2022}
Krasnokutski, S.~A., Chuang, K.-J., Jäger, C., Ueberschaar, N., \& Henning, T.
  2022, Nat Astron, 1, \dodoi{10.1038/s41550-021-01577-9}

\bibitem[{Krasnokutski {et~al.}(2020)Krasnokutski, Jäger, \&
  Henning}]{krasnokutski_condensation_2020}
Krasnokutski, S.~A., Jäger, C., \& Henning, T. 2020, ApJ, 889, 67,
  \dodoi{10.3847/1538-4357/ab60a1}

\bibitem[{Krasnokutski {et~al.}(2017)Krasnokutski, Goulart, Gordon, Ritsch,
  Jäger, Rastogi, Salvenmoser, Henning, \&
  Scheier}]{krasnokutski_low-temperature_2017}
Krasnokutski, S.~A., Goulart, M., Gordon, E.~B., {et~al.} 2017, \apj, 847, 89,
  \dodoi{10.3847/1538-4357/aa88a4}

\bibitem[{Kruse \& Grimme(2012)}]{kruse_geometrical_2012}
Kruse, H., \& Grimme, S. 2012, J. Chem. Phys., 136, 154101,
  \dodoi{10.1063/1.3700154}

\bibitem[{Kuwahata {et~al.}(2015)Kuwahata, Hama, Kouchi, \&
  Watanabe}]{Kuwahata2015}
Kuwahata, K., Hama, T., Kouchi, A., \& Watanabe, N. 2015, Phys. Rev. Lett.,
  115, 133201, \dodoi{10.1103/PhysRevLett.115.133201}

\bibitem[{{Lamberts} {et~al.}(2022){Lamberts}, {Fedoseev}, {van Hemert},
  {Qasim}, {Chuang}, {Santos}, \& {Linnartz}}]{Lamberts2022_methane}
{Lamberts}, T., {Fedoseev}, G., {van Hemert}, M.~C., {et~al.} 2022, \apj, 928,
  48, \dodoi{10.3847/1538-4357/ac51d1}

\bibitem[{{Lefloch} {et~al.}(2017){Lefloch}, {Ceccarelli}, {Codella}, {Favre},
  {Podio}, {Vastel}, {Viti}, \& {Bachiller}}]{Lefloch2017}
{Lefloch}, B., {Ceccarelli}, C., {Codella}, C., {et~al.} 2017, \mnras, 469,
  L73, \dodoi{10.1093/mnrasl/slx050}

\bibitem[{Liu \& McLean(1973)}]{liu_accurate_1973}
Liu, B., \& McLean, A.~D. 1973, J. Chem. Phys., 59, 4557,
  \dodoi{10.1063/1.1680654}

\bibitem[{{Mart{\'\i}n} {et~al.}(2021){Mart{\'\i}n}, {Mangum}, {Harada},
  {Costagliola}, {Sakamoto}, {Muller}, {Aladro}, {Tanaka}, {Yoshimura},
  {Nakanishi}, {Herrero-Illana}, {M{\"u}hle}, {Aalto}, {Behrens}, {Colzi},
  {Emig}, {Fuller}, {Garc{\'\i}a-Burillo}, {Greve}, {Henkel}, {Holdship},
  {Humire}, {Hunt}, {Izumi}, {Kohno}, {K{\"o}nig}, {Meier}, {Nakajima},
  {Nishimura}, {Padovani}, {Rivilla}, {Takano}, {van der Werf}, {Viti}, \&
  {Yan}}]{Martin2021-extragal}
{Mart{\'\i}n}, S., {Mangum}, J.~G., {Harada}, N., {et~al.} 2021, \aap, 656,
  A46, \dodoi{10.1051/0004-6361/202141567}

\bibitem[{{McClure} {et~al.}(2023){McClure}, {Rocha}, {Pontoppidan}, {Crouzet},
  {Chu}, {Dartois}, {Lamberts}, {Noble}, {Pendleton}, {Perotti}, {Qasim},
  {Rachid}, {Smith}, {Sun}, {Beck}, {Boogert}, {Brown}, {Caselli}, {Charnley},
  {Cuppen}, {Dickinson}, {Drozdovskaya}, {Egami}, {Erkal}, {Fraser}, {Garrod},
  {Harsono}, {Ioppolo}, {Jim{\'e}nez-Serra}, {Jin}, {J{\o}rgensen},
  {Kristensen}, {Lis}, {McCoustra}, {McGuire}, {Melnick}, {{\"O}berg},
  {Palumbo}, {Shimonishi}, {Sturm}, {van Dishoeck}, \&
  {Linnartz}}]{McClure2023}
{McClure}, M.~K., {Rocha}, W.~R.~M., {Pontoppidan}, K.~M., {et~al.} 2023,
  Nature Astronomy, \dodoi{10.1038/s41550-022-01875-w}

\bibitem[{{McGuire}(2022)}]{mcguire2021}
{McGuire}, B.~A. 2022, \apjs, 259, 30, \dodoi{10.3847/1538-4365/ac2a48}

\bibitem[{Meisner \& K{\"a}stner(2016)}]{meisner2016atom}
Meisner, J., \& K{\"a}stner, J. 2016, Angew. Chem. Int. Ed., 55, 5400,
  \dodoi{10.1002/anie.201511028}

\bibitem[{Meisner {et~al.}(2017)Meisner, Lamberts, \&
  Kästner}]{meisner_lambert_kast}
Meisner, J., Lamberts, T., \& Kästner, J. 2017, ACS Earth and Space Chemistry,
  1, 399, \dodoi{10.1021/acsearthspacechem.7b00052}

\bibitem[{{Miksch} {et~al.}(2021){Miksch}, {Riffelt}, {Oliveira},
  {K{\"a}stner}, \& {Molpeceres}}]{Molp_tun_addition}
{Miksch}, A.~M., {Riffelt}, A., {Oliveira}, R., {K{\"a}stner}, J., \&
  {Molpeceres}, G. 2021, \mnras, 505, 3157, \dodoi{10.1093/mnras/stab1514}

\bibitem[{{Molpeceres} \& {K{\"a}stner}(2021)}]{Molp_tun_phosph}
{Molpeceres}, G., \& {K{\"a}stner}, J. 2021, \apj, 910, 55,
  \dodoi{10.3847/1538-4357/abe38c}

\bibitem[{Molpeceres {et~al.}(2021)Molpeceres, Kästner, Fedoseev, Qasim,
  Schömig, Linnartz, \& Lamberts}]{molpeceres_carbon_2021}
Molpeceres, G., Kästner, J., Fedoseev, G., {et~al.} 2021, J. Phys. Chem.
  Lett., 12, 10854, \dodoi{10.1021/acs.jpclett.1c02760}

\bibitem[{{Molpeceres} \& {Rivilla}(2022)}]{molp_tun_rivill}
{Molpeceres}, G., \& {Rivilla}, V.~M. 2022, \aap, 665, A27,
  \dodoi{10.1051/0004-6361/202243892}

\bibitem[{{Molpeceres} {et~al.}(2023{\natexlab{a}}){Molpeceres}, {Rivilla},
  {Furuya}, {K{\"a}stner}, {Mat{\'e}}, \& {Aikawa}}]{Molp_tun_nh2oh}
{Molpeceres}, G., {Rivilla}, V.~M., {Furuya}, K., {et~al.} 2023{\natexlab{a}},
  \mnras, 521, 6061, \dodoi{10.1093/mnras/stad892}

\bibitem[{{Molpeceres} {et~al.}(2023{\natexlab{b}}){Molpeceres}, {Zaverkin},
  {Furuya}, {Aikawa}, \& {K{\"a}stner}}]{Molpeceres2023}
{Molpeceres}, G., {Zaverkin}, V., {Furuya}, K., {Aikawa}, Y., \& {K{\"a}stner},
  J. 2023{\natexlab{b}}, \aap, 673, A51, \dodoi{10.1051/0004-6361/202346073}

\bibitem[{{Molpeceres} {et~al.}(2022){Molpeceres}, {Jim{\'e}nez-Serra}, {Oba},
  {Nguyen}, {Watanabe}, {de la Concepci{\'o}n}, {Mat{\'e}}, {Oliveira}, \&
  {K{\"a}stner}}]{molp_tun_abstractions}
{Molpeceres}, G., {Jim{\'e}nez-Serra}, I., {Oba}, Y., {et~al.} 2022, \aap, 663,
  A41, \dodoi{10.1051/0004-6361/202243366}

\bibitem[{{Muller} {et~al.}(2011){Muller}, {Beelen}, {Gu{\'e}lin}, {Aalto},
  {Black}, {Combes}, {Curran}, {Theule}, \& {Longmore}}]{Muller2011-extragal}
{Muller}, S., {Beelen}, A., {Gu{\'e}lin}, M., {et~al.} 2011, \aap, 535, A103,
  \dodoi{10.1051/0004-6361/201117096}

\bibitem[{Neese {et~al.}(2020)Neese, Wennmohs, Becker, \&
  Riplinger}]{neese_orca_2020}
Neese, F., Wennmohs, F., Becker, U., \& Riplinger, C. 2020, J. Chem. Phys.,
  152, 224108, \dodoi{10.1063/5.0004608}

\bibitem[{{{\"O}berg} {et~al.}(2015){{\"O}berg}, {Guzm{\'a}n}, {Furuya}, {Qi},
  {Aikawa}, {Andrews}, {Loomis}, \& {Wilner}}]{Oberg2015}
{{\"O}berg}, K.~I., {Guzm{\'a}n}, V.~V., {Furuya}, K., {et~al.} 2015, \nat,
  520, 198, \dodoi{10.1038/nature14276}

\bibitem[{{Palumbo} {et~al.}(1999){Palumbo}, {Castorina}, \&
  {Strazzulla}}]{Palumbo1999A&A...342..551P}
{Palumbo}, M.~E., {Castorina}, A.~C., \& {Strazzulla}, G. 1999, \aap, 342, 551

\bibitem[{Pantaleone {et~al.}(2021)Pantaleone, Enrique-Romero, Ceccarelli,
  Ferrero, Balucani, Rimola, \& Ugliengo}]{pantaleone_h2_2021}
Pantaleone, S., Enrique-Romero, J., Ceccarelli, C., {et~al.} 2021, ApJ, 917,
  49, \dodoi{10.3847/1538-4357/ac0142}

\bibitem[{Pantaleone {et~al.}(2020)Pantaleone, Enrique-Romero, Ceccarelli,
  Ugliengo, Balucani, \& Rimola}]{pantaleone_chemical_2020}
---. 2020, ApJ, 897, 56, \dodoi{10.3847/1538-4357/ab8a4b}

\bibitem[{Perrero {et~al.}(2022)Perrero, Enrique-Romero, Martínez-Bachs,
  Ceccarelli, Balucani, Ugliengo, \& Rimola}]{perrero_non-energetic_2022}
Perrero, J., Enrique-Romero, J., Martínez-Bachs, B., {et~al.} 2022, ACS Earth
  Space Chem., 6, 496, \dodoi{10.1021/acsearthspacechem.1c00369}

\bibitem[{Pezzella \& Meuwly(2019)}]{pezzella_o2_2019}
Pezzella, M., \& Meuwly, M. 2019, Phys. Chem. Chem. Phys., 21, 6247,
  \dodoi{10.1039/C8CP07474G}

\bibitem[{{Pickett} {et~al.}(1998){Pickett}, {Poynter}, {Cohen}, {Delitsky},
  {Pearson}, \& {M{\"u}ller}}]{Pickett1998-JPL}
{Pickett}, H.~M., {Poynter}, R.~L., {Cohen}, E.~A., {et~al.} 1998, \jqsrt, 60,
  883, \dodoi{10.1016/S0022-4073(98)00091-0}

\bibitem[{Potapov {et~al.}(2021)Potapov, Krasnokutski, Jäger, \&
  Henning}]{potapov_new_2021}
Potapov, A., Krasnokutski, S.~A., Jäger, C., \& Henning, T. 2021, ApJ, 920,
  111, \dodoi{10.3847/1538-4357/ac1a70}

\bibitem[{{Qasim} {et~al.}(2020){Qasim}, {Fedoseev}, {Chuang}, {He}, {Ioppolo},
  {van Dishoeck}, \& {Linnartz}}]{QASIM_CH4}
{Qasim}, D., {Fedoseev}, G., {Chuang}, K.~J., {et~al.} 2020, Nature Astronomy,
  4, 781, \dodoi{10.1038/s41550-020-1054-y}

\bibitem[{Rimola {et~al.}(2021)Rimola, Ceccarelli, Balucani, \&
  Ugliengo}]{rimola_interaction_2021}
Rimola, A., Ceccarelli, C., Balucani, N., \& Ugliengo, P. 2021, Front. Astron.
  Space Sci., 8.
\newblock \url{https://www.frontiersin.org/article/10.3389/fspas.2021.655405}

\bibitem[{Rimola {et~al.}(2018)Rimola, Skouteris, Balucani, Ceccarelli,
  Enrique-Romero, Taquet, \& Ugliengo}]{rimola_can_2018}
Rimola, A., Skouteris, D., Balucani, N., {et~al.} 2018, ACS Earth Space Chem.,
  2, 720, \dodoi{10.1021/acsearthspacechem.7b00156}

\bibitem[{Rimola {et~al.}(2010)Rimola, Sodupe, \& Ugliengo}]{rimola2010deep}
Rimola, A., Sodupe, M., \& Ugliengo, P. 2010, Phys. Chem. Chem. Phys., 12,
  5285, \dodoi{10.1039/B923439J}

\bibitem[{{Rubin} {et~al.}(1971){Rubin}, {Swenson}, {Benson}, {Tigelaar}, \&
  {Flygare}}]{Rubin1971-formamide}
{Rubin}, R.~H., {Swenson}, G.~W., J., {Benson}, R.~C., {Tigelaar}, H.~L., \&
  {Flygare}, W.~H. 1971, \apjl, 169, L39, \dodoi{10.1086/180810}

\bibitem[{{Schutte} {et~al.}(1992){Schutte}, {Allamandola}, \&
  {Sandford}}]{Schutte1992AdSpR..12d..47S}
{Schutte}, W.~A., {Allamandola}, L.~J., \& {Sandford}, S.~A. 1992, Adv. Space
  Res., 12, 47, \dodoi{10.1016/0273-1177(92)90152-N}

\bibitem[{{Scibelli} \& {Shirley}(2020)}]{Scibelli2020-PSC}
{Scibelli}, S., \& {Shirley}, Y. 2020, \apj, 891, 73,
  \dodoi{10.3847/1538-4357/ab7375}

\bibitem[{{Senevirathne} {et~al.}(2017){Senevirathne}, {Andersson}, {Dulieu},
  \& {Nyman}}]{Senevirathne2017-Hdiffusion}
{Senevirathne}, B., {Andersson}, S., {Dulieu}, F., \& {Nyman}, G. 2017, Mol.
  Astrophys., 6, 59, \dodoi{10.1016/j.molap.2017.01.005}

\bibitem[{Shimonishi {et~al.}(2018)Shimonishi, Nakatani, Furuya, \&
  Hama}]{shimonishi_adsorption_2018}
Shimonishi, T., Nakatani, N., Furuya, K., \& Hama, T. 2018, ApJ, 855, 27,
  \dodoi{10.3847/1538-4357/aaaa6a}

\bibitem[{{Skouteris} {et~al.}(2018){Skouteris}, {Balucani}, {Ceccarelli},
  {Vazart}, {Puzzarini}, {Barone}, {Codella}, \&
  {Lefloch}}]{Skouteris2018-ethanoltree}
{Skouteris}, D., {Balucani}, N., {Ceccarelli}, C., {et~al.} 2018, \apj, 854,
  135, \dodoi{10.3847/1538-4357/aaa41e}

\bibitem[{{Skouteris} {et~al.}(2017){Skouteris}, {Vazart}, {Ceccarelli},
  {Balucani}, {Puzzarini}, \& {Barone}}]{Skouteris2017-formamide}
{Skouteris}, D., {Vazart}, F., {Ceccarelli}, C., {et~al.} 2017, \mnras, 468,
  L1, \dodoi{10.1093/mnrasl/slx012}

\bibitem[{{Strazzulla}(1997)}]{Strazzulla1997AdSpR..19.1077S}
{Strazzulla}, G. 1997, Adv. Space Res., 19, 1077,
  \dodoi{10.1016/S0273-1177(97)00356-6}

\bibitem[{{Suzuki} {et~al.}(2018){Suzuki}, {Majumdar}, {Ohishi}, {Saito},
  {Hirota}, \& {Wakelam}}]{Suzuki2018}
{Suzuki}, T., {Majumdar}, L., {Ohishi}, M., {et~al.} 2018, \apj, 863, 51,
  \dodoi{10.3847/1538-4357/aad087}

\bibitem[{{Taquet} {et~al.}(2012){Taquet}, {Ceccarelli}, \&
  {Kahane}}]{Taquet2012-grainoble}
{Taquet}, V., {Ceccarelli}, C., \& {Kahane}, C. 2012, \aap, 538, A42,
  \dodoi{10.1051/0004-6361/201117802}

\bibitem[{{Tielens} \& {Hagen}(1982)}]{Tielens1982}
{Tielens}, A.~G.~G.~M., \& {Hagen}, W. 1982, \aap, 114, 245

\bibitem[{{Tsuge} {et~al.}(2023){Tsuge}, {Molpeceres}, {Aikawa}, \&
  {Watanabe}}]{tsuge2023surface}
{Tsuge}, M., {Molpeceres}, G., {Aikawa}, Y., \& {Watanabe}, N. 2023, Nature
  Astronomy, 1, \dodoi{https://doi.org/10.1038/s41550-023-02071-0}

\bibitem[{{Vastel} {et~al.}(2014){Vastel}, {Ceccarelli}, {Lefloch}, \&
  {Bachiller}}]{Vastel2014-PSCiCOM}
{Vastel}, C., {Ceccarelli}, C., {Lefloch}, B., \& {Bachiller}, R. 2014, \apjl,
  795, L2, \dodoi{10.1088/2041-8205/795/1/L2}

\bibitem[{Vasyunin \& Herbst(2013)}]{vasyunin_reactive_2013}
Vasyunin, A.~I., \& Herbst, E. 2013, ApJ, 769, 34,
  \dodoi{10.1088/0004-637X/769/1/34}

\bibitem[{{Vazart} {et~al.}(2020){Vazart}, {Ceccarelli}, {Balucani}, {Bianchi},
  \& {Skouteris}}]{Vazart2020-acetaldehyde}
{Vazart}, F., {Ceccarelli}, C., {Balucani}, N., {Bianchi}, E., \& {Skouteris},
  D. 2020, \mnras, 499, 5547, \dodoi{10.1093/mnras/staa3060}

\bibitem[{Vydrov \& Van~Voorhis(2010)}]{vydrov_nonlocal_2010}
Vydrov, O.~A., \& Van~Voorhis, T. 2010, J. Chem. Phys., 133, 244103,
  \dodoi{10.1063/1.3521275}

\bibitem[{{Wakelam} {et~al.}(2021){Wakelam}, {Dartois}, {Chabot}, {Spezzano},
  {Navarro-Almaida}, {Loison}, \& {Fuente}}]{Wakelam2021-CRdes}
{Wakelam}, V., {Dartois}, E., {Chabot}, M., {et~al.} 2021, \aap, 652, A63,
  \dodoi{10.1051/0004-6361/202039855}

\bibitem[{{Walsh} {et~al.}(2016){Walsh}, {Loomis}, {{\"O}berg}, {Kama}, {van 't
  Hoff}, {Millar}, {Aikawa}, {Herbst}, {Widicus Weaver}, \&
  {Nomura}}]{Walsh2016-ch3oh}
{Walsh}, C., {Loomis}, R.~A., {{\"O}berg}, K.~I., {et~al.} 2016, \apjl, 823,
  L10, \dodoi{10.3847/2041-8205/823/1/L10}

\bibitem[{Watanabe \& Kouchi(2008)}]{watanabe_ice_2008}
Watanabe, N., \& Kouchi, A. 2008, Progress in Surface Science, 83, 439,
  \dodoi{10.1016/j.progsurf.2008.10.001}

\bibitem[{Weigend \& Ahlrichs(2005)}]{weigend_balanced_2005}
Weigend, F., \& Ahlrichs, R. 2005, Phys. Chem. Chem. Phys., 7, 3297,
  \dodoi{10.1039/B508541A}

\bibitem[{Woon(2021)}]{woon_quantum_2021}
Woon, D.~E. 2021, Acc. Chem. Res., 54, 490,
  \dodoi{10.1021/acs.accounts.0c00717}

\bibitem[{{Yang} {et~al.}(2022){Yang}, {Green}, {Pontoppidan}, {Bergner},
  {Cleeves}, {Evans}, {Garrod}, {Jin}, {Kim}, {Kim}, {Lee}, {Sakai},
  {Shingledecker}, {Shope}, {Tobin}, \& {van Dishoeck}}]{Yang2022-JWST}
{Yang}, Y.-L., {Green}, J.~D., {Pontoppidan}, K.~M., {et~al.} 2022, \apjl, 941,
  L13, \dodoi{10.3847/2041-8213/aca289}

\bibitem[{Zamirri {et~al.}(2019)Zamirri, Ugliengo, Ceccarelli, \&
  Rimola}]{zamirri_quantum_2019}
Zamirri, L., Ugliengo, P., Ceccarelli, C., \& Rimola, A. 2019, ACS Earth Space
  Chem., 3, 1499, \dodoi{10.1021/acsearthspacechem.9b00082}

\bibitem[{Ásgeirsson {et~al.}(2021)Ásgeirsson, Birgisson, Bjornsson, Becker,
  Neese, Riplinger, \& Jónsson}]{asgeirsson_nudged_2021}
Ásgeirsson, V., Birgisson, B.~O., Bjornsson, R., {et~al.} 2021, J. Chem.
  Theory Comput., 17, 4929, \dodoi{10.1021/acs.jctc.1c00462}

\bibitem[{Öberg(2016)}]{Oberg2016}
Öberg, K.~I. 2016, Chem. Rev., 116, 9631, \dodoi{10.1021/acs.chemrev.5b00694}

\end{thebibliography}
\bibliographystyle{aasjournal} 

\end{document}